\documentclass[sigconf, xcolor=table]{acmart}
\usepackage{balance}
\usepackage{url}
\usepackage{amsmath,amssymb,amsfonts}
\usepackage{multirow}
\usepackage{graphicx}
\usepackage{textcomp}
\usepackage{listings}
\usepackage{booktabs}
\usepackage{color}
\usepackage{threeparttable}
\usepackage{pdfpages}  
\usepackage[noend]{algpseudocode}
\usepackage[ruled]{algorithm2e}

\def\BibTeX{{\rm B\kern-.05em{\sc i\kern-.025em b}\kern-.08emT\kern-.1667em\lower.7ex\hbox{E}\kern-.125emX}}

\setcopyright{none} % No copyright notice required for submissions
\renewcommand\footnotetextcopyrightpermission[1]{} % removes footnote with conference information in first column
\begin{document}
\pagestyle{plain}
%
% The "title" command has an optional parameter, allowing the author to define a "short title" to be used in page headers.
\title{Automated Deobfuscation of Android Native Binary Code}
\newcommand{\Name}{DiANa\space}

\author{Zeliang Kan, Haoyu Wang}
\affiliation{Beijing University of Posts and Telecommunications}

\author{Lei Wu}
\affiliation{Zhejiang University}

\author{Yao Guo}
\affiliation{Peking University}

\author{Daniel Xiapu Luo}
\affiliation{The Hong Kong Polytechnic University}

\begin{abstract}
With the popularity of Android apps, different techniques have been proposed to enhance app protection. As an effective approach to prevent reverse engineering, obfuscation can be used to serve both benign and malicious purposes. In recent years, more and more sensitive logic or data have been implemented as obfuscated native code because of the limitations of Java bytecode. As a result, native code obfuscation becomes a great obstacle for security analysis to understand the complicated logic. In this paper, we propose DiANa, an automated system to facilitate the deobfuscation of native binary code in Android apps. Specifically, given a binary obfuscated by Obfuscator-LLVM (the most popular native code obfuscator), DiANa is capable of recovering the original Control Flow Graph. To the best of our knowledge, DiANa is the first system that aims to tackle the problem of Android native binary deobfuscation. We have applied DiANa in different scenarios, and the experimental results demonstrate the effectiveness of DiANa based on generic similarity comparison metrics.
\end{abstract}

\maketitle

\section{Introduction}

% Para1: Android apps grow explosively, could be developed in both Java and C++
% Android applications (apps) have seen wide spread adoption recent years. The number of apps in Google Play has surpassed 3.5 Million mark by the end of April 2018~\cite{AppNumber}. Android PacKage (APK) is the package file format used by the Android Operating System for distribution and installation of mobile apps. APKs are generally written in Java, while Android also supports incorporating native code binaries (written in C/C++) as part of the apps to boost app developing process. 

%% native code
Android apps have received widespread adoption recently \cite{AppNumber}. Besides the standard Android programming model in Java, Android NDK \cite{NDK} was introduced to allow developers to include native code binaries (write in C/C++) in their apps.

\begin{figure}[t]
\centerline{\includegraphics[width=0.48\textwidth]{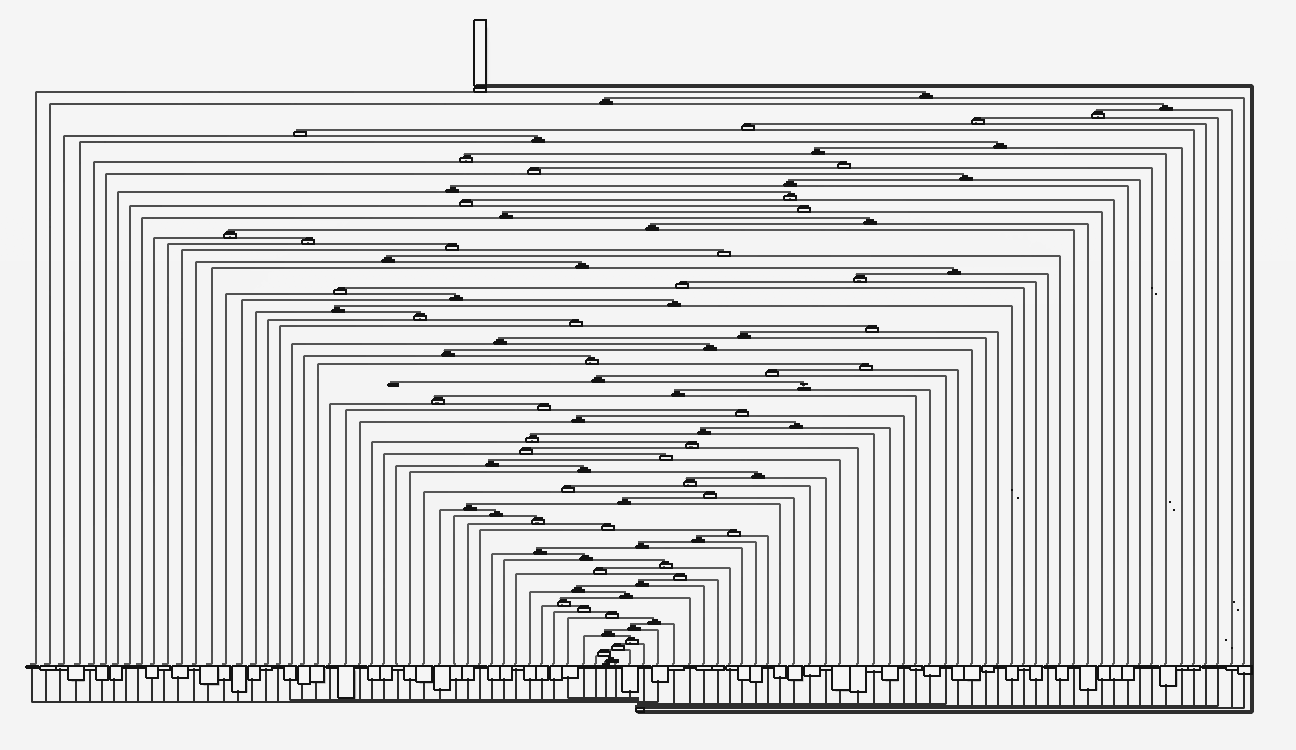}}
%%\vspace{-0.1in}
\caption{The CFG of an obfuscation function in POC of the TowelRoot exploit.}
%\vspace{-0.2in}
\label{fig:motivating}
\end{figure}

Recent work suggested that native code had been widely used~\cite{sun2014nativeguard, alam2016droidnative, afonso2016going} in Android apps, which severely complicates the process of static analysis. As Java bytecode can be easily decompiled, malware developers usually hide the malicious payload and core functionalities in the native code to evade detection~\cite{malware1, malware2, malware3, afonso2016going}. 
Even worse, native code can also be obfuscated, which further increases the difficulty of security analysis. For example, the towelroot exploit (CVE-2014-3153)~\cite{Towelroot1, CVE2014}, one of the biggest Root Exploits family in Android, was found obfuscated at the native code level by O-LLVM. It took a lot of efforts for security researchers to dive into the technical details of the code due to obfuscation. Figure~\ref{fig:motivating} shows the control flow graph of an obfuscated function ``search\_goodnum'', which is one of the key components of a Proof-of-Concept (PoC) implementation of the towelroot exploit. Apparently, the obfuscated control flow becomes too complex to understand the real logic. 

Although several attempts~\cite{Bichsel2016Statistical, Baumann2017Anti} have discussed the deobfuscation of Android apps, all of them have been focused on Java-level deobfuscation.
% (either Java source code or bytecode). 
For example, DeGuard~\cite{Bichsel2016Statistical} was proposed to reverse layout obfuscation (naming obfuscation) generated by ProGuard. Their key idea is to learn a probabilistic model over a large number of non-obfuscated apps and to use this model to deobfuscate new APKs. Layout obfuscation is the easiest one in Android app obfuscation, which does not alter the program logic (e.g., control flow) of the apps. To the best of our knowledge, no previous studies have attempted to tackle native code deobfuscation for Android yet. 
% \YG{How about other work besides DeGuard??}

\textbf{Motivation.} The focus of this paper is deobfuscating the native code in Android apps. To be specific, we focus on the deobfuscation of native code that was protected by Obfuscator-LLVM (O-LLVM), which is the most popular native code obfuscator. O-LLVM is a set of obfuscating code transformations suite, which is implemented as middle-end passes in the LLVM compilation flow. O-LLVM offers three obfuscating methods: Instruction Substitution (InsSub), Bogus Control Flow (BCF) and Control Flow Flattening (CFF). Many popular packing tools, e.g., Baidu Jiagu~\cite{baidujiagu} and Bancle~\cite{Bangcle}, are implemented based on O-LLVM. 
% \haoyu{One or two sentences here to briefly describe OLLVM.---done} 
% \haoyu{Briefly describe two related work here.}

\textbf{Challenges.} Binary code deobfuscation is not a new task introduced in Android, as many studies have already been proposed for PC platforms (e.g., x86). Udupa \textit{et al.}~\cite{udupa2005deobfuscation} use static techniques to remove the dead edges added by the obfuscator. Yadegari \textit{et al.}~\cite{yadegari2015generic} proposed a deobfuscating approach based on the execution trace of the executable code. 
However, it is non-trivial to deobfuscate Android native binary code because there are no existing deobfuscation tools that can deal with full O-LLVM obfuscation, especially in the scenario of Android apps and for the ARM platform. Specifically, we face the following main challenges in this work:
\begin{itemize}
\item 
\textbf{Complexity due to instruction optimization on ARM}. Due to the specific optimizations on the ARM platform, some obfuscated operations may not be independent. It may be interleaved with normal instructions and operations. 
Thus it becomes a challenge to comprehensively and accurately pinpoint the obfuscated instructions. 
% 由于arm平台的优化，导致混淆的部分与原始信息的混合复用，如何在这种情况下将两者区别开来，并恢复混淆部分，成为了我们首先面对的问题。

% \haoyu{rewrite}
\item 
\textbf{Difficulty in control flow recovery due to basic block splitting}.
%  and basic block splitting 
% 由于ollvm的混淆使用分发器对原始基本块进行调度，在这种情况下上下文的继承关系就显得特别重要，如何在尽可能多的继承上下文信息，并不造成路径爆炸，也成为了一个关键的challenge
Control Flow Flattening in O-LLVM obfuscation can completely destroy the original control flow. At the same time, original basic blocks can also be split or partially merged, which may cause the separation of original semantic information. It becomes difficult to properly recover control flow in this kind of situations.
% OLLVM可以将原始控制流摧毁，并且原始的基本块在混淆程序中会被拆分和部分复用，从而造成携带信息与原始信息的差异。如何在这种情况下正确的恢复控制流。
% 由于OLLVM的反混淆工作是flow-sensitive的，在符号执行时，由于路径约束的不确定性，直接的符号执行会造成路径爆炸的问题，如何尽可能多的保留上下文信息，并不造成路径的爆炸，也成为了一个关键的challenge
\item 
% \textbf{How to perform flow-sensitive symbolic execution}.\haoyu{rewrite}
\textbf{Challenges in symbolic execution due to unknown path constrains}.
We will apply symbolic execution as one of the key steps in deobfuscation for the purpose of systematically exploring multiple program paths, similar as in previous work~\cite{guillot2010automatic,xu2017cryptographic,gabriel2014deobfuscation}. However, since the recovering process of O-LLVM obfuscated binary is flow-sensitive, incomplete context information may lead to incomplete or wrong path exploration,
while performing symbolic execution directly will cause the path explosion problem. Thus we need to find a way to retain the context information as much as possible, while avoiding path explosion during symbolic execution.

% How to maximize the context inheritance during the process of recovering the original control flow becomes a key challenge of our work.
% \footnote{For example, the constraint and calculation in a basic block may be separated into two blocks. It can only obtain incomplete information when analyze a }

% Thus, the context inheritance become particularly important in the recover process. How to inherit the context information as much as possible and do not cause the path explosion also became a key challenge.
% \haoyu{A short title here to name the challenge.}

% \item \textbf{Hybrid Obfuscation Techniques.} In general, malware developers could take advantage of hybrid obfuscation techniques to generate a robust malware sample. As O-LLVM provided different techniques to obfuscate code in both instructions and control flow aspect, how to identify what obfuscation techniques are used, and perform the deobfuscation actions at hybrid cases (rather than obfuscation-specific approach), has remained as a challenge.
% (\haoyu{briefly describe what is pass here}) 
% 对于ollvm多种情况混合的时候，如何进行特征判定，进而进行后续的恢复工作也成为了我们工作中遇到的问题
% For the unknown of the analysis object in reality situation, how to judge the obfuscation approach of OLLVM, and then carry out the subsequent recovery work, has become a challenge in our work too.
\end{itemize}

\textbf{Approach}
In this paper, we propose DiANa (\underline{\textbf{D}}eobfusca-t\underline{\textbf{i}}on of \underline{\textbf{A}}ndroid \underline{\textbf{Na}}tive Binary Code), a new approach for automated deobfuscation of Android native binary code. 
% It uses the method of taint analysis and symbolic execution to recover the O-LLVM obfuscated Android native functions.
Technically, \Name works by combining static taint analysis and symbolic execution to remove the obfuscation introduced by different obfuscating techniques in O-LLVM. One key feature of \Name is that we introduce \textit{taint analysis} to perform semantic level deobfuscation, while considering both general and compiler optimization situation. We also exploit flow-sensitive \textit{symbolic execution} to rebuild the seriously obfuscated control flow. 
To overcome the challenge of basic block splitting, we chop the original control flow, select analysis targets through static features, and dynamically adjust the analysis target sequence to maximize context inheritance.

% \YG{More on evaluation results HERE.}
% 为了克服基本块拆分带来的挑战，我们将原始控制流拆分，通过静态特征选择分析目标，并动态调整分析目标序列，以最大程度的继承上下文

% \haoyu{describe more about our system and how you overcome the challenges.}
% It is noticeable that DiANA also can detect what kinds of obfuscating approach are used and makes deobfuscation based on the 
% \haoyu{INTRODUCE SOME KEY TECHNIQUES HERE. How to solve the aforementioned challenges?}
% \begin{itemize}
% \item Approach1
% % 对ARM平台的混淆特征进行了全面的总结，包括多种pass进行混淆的情况
% \item Approach2
% % 对于OLLVM的部分混淆使用污点分析，将疑似混淆操作的寄存器置为taint，以应对arm的指令优化，确保反混淆的完整性
% \item Approach3
% % 对符号执行进行CFF反混淆的时候，对基本块序列的执行序列进行动态调整，以最大程度上的继承上下文
% \end{itemize}

Experiments on multiple benchmarks and real-world cases suggest that our approach can accurately deobfuscate native code obfuscated by O-LLVM. We believe \Name can be used by security analysts to make it easier to inspect native code, even if it was heavily obfuscated.
% In addition, evaluation proves that \Name works well with the hybrid obfuscation due to the comprehensive feature summary.
This paper makes the following major contributions:
\begin{itemize}
\item To the best of our knowledge, this is the first work to tackle native code deobfuscation in Android apps. Our approach is able to handle different obfuscations provided by O-LLVM, the most popular obfuscator for Android. 

\item We design and implement DiANa, a system that can successfully deobfuscate native code in Android apps. \Name leverages taint analysis to overcome the complication due to ARM-specific optimizations. It also deploys flow-sensitive symbolic execution to tackle  Control Flow Flattening obfuscation in O-LLVM. 
%   We have a set of optimization methods to both the deobfuscating process and the final output. These can not only improve the efficiency of deobfuscation work, but also make it easier to understand recovered results.
% We believe our research could help boost the native-level malware detection and analysis.
% \haoyu{need a contribution here}\textbf{\haoyu{suggest write the optimization/novel aspects of our work here}} 

\item Our evaluation on a set of Android apps demonstrates that \Name is effective in native code deobfuscation and could become a valuable tool for tasks including program analysis and malware detection.  
%We have implemented a deobfuscation system DiANa, and applied it to various benchmarks and real-world exploits to evaluate the effectiveness of our system. 
We will release our system and the benchmarks to the research community at: 
\url{https://github.com/DiANa-deobfuscation-2020/DiANa}
%%\footnote{The code of symbol execution in CFF is inspried by the deflat script of the GitHub user liumengdeqq on x86 platform.}
\end{itemize}
\section{Background and Related Work}
\label{sec:background}

To the best of our knowledge, this is the first work that tackles Android native code deobfuscation in our community. Nevertheless, there are already many studies on code obfuscation and deobfuscation for applications on both mobile and PC platforms.

\subsection{Android App Obfuscation}

To increase the bar for reverse engineering and hinder cracking, and prevent theft of intellectual property, code obfuscation techniques have been widely used in Android apps~\cite{dong2018understanding}, for both legitimate apps and malware.
Many tools allow developers to obfuscate their apps~\cite{proguard, DashO, DexGuard, DexProtector, allatori, junod2015obfuscator}. For example, ProGuard is a popular obfuscation tool integrated in the Android build system. These tools may operate at different levels, e.g., Java code level, Dex bytecode level and native code level.

Native code level protection is much stronger than Java-level protection, thus state-of-the-art commercial packers utilize native code obfuscation to increase the complexity of reverse engineering~\cite{duan2018things}. As a side effect, Android malware also take advantage of native code obfuscation to evade detection.

\textbf{Obfuscator-LLVM (O-LLVM)} is one of the most widely used code obfuscator for both x86 and ARM platforms~\cite{junod2015obfuscator}.
It is implemented as middle-end passes in the LLVM compilation process, which offers guaranteed compatibility with LLVM. 
% It was first introduced by Junod, P., et al.\cite{junod2015obfuscator} and open-sourced on Git-Hub\cite{githubollvm}. 
% \haoyu{should describe clear what is pass here.} 
It offers the following three obfuscation techniques: 
% \emph{Instruction Substitution}, \emph{Bogus Control Flow} and \emph{Control Flow Flattening}. 

\emph{\textbf{(1) Instruction Substitution (InsSub)}}.
InsSub is the simplest obfuscation technique in O-LLVM. It replaces the standard binary operations, e.g., arithmetic (ADD, SUB) or boolean ones (AND, OR and XOR), by functionally equivalent but more complicated sequences of instructions. 
For each kind of operations, there are multiple ways to perform obfuscation. InsSub chooses one at random to increase code diversification in the resulting obfuscated code. 

\emph{\textbf{(2) Bogus Control Flow (BCF)}.}
BCF modifies the CFG by adding a conditional jump to randomly selected basic blocks, which either points to the original basic block or to a fake basic block looping back to the conditional jump block\cite{junod2015obfuscator}. In order to evade detection and optimization by the optimizers, an opaque predicate~\cite{wiki:opaque} (i.e., a mathematical expression which always evaluating to the same value) is used to ensure that only the original basic block is executed during run-time. 

\emph{\textbf{(3) Control Flow Flattening (CFF)}.}
This is the most effective, and the most difficult to deobfuscate, pass in O-LLVM. The basic idea is to remove all easily identifiable conditional jumps (e.g., \texttt{IF-ELSE}) and looping structures (e.g., \texttt{WHILE,FOR}), and use a big switch construct to route the code control flow through the proper basic blocks. The flow dispatcher chooses the next block using a routing variable, which resets in each basic block and leads the flow to the next correct basic block. Besides, there is a compilation flag that enables further breaking the code structure by artificially increasing the number of basic blocks in a function.

In general, InsSub and BCF are both semantic obfuscations that work at the instruction level, while CFF is an overall remodeling of the control flow in a function.

\vspace{-0.1in}
\subsection{Mobile App Deobfuscation}

Although several attempts have been focused on the deobfuscation of Android apps, all of them deal with Java-level deobfuscation. DeGuard \cite{Bichsel2016Statistical} was proposed to deal with layout obfuscation introduced by ProGuard~\cite{proguard}. The key idea is to summarize a probabilistic model by learning unobfuscated apps on a large scale and then use the model to recover the obfuscated code. Also, Baumann \textit{et al.}~\cite{Baumann2017Anti} use a similar approach to perform ProGuard deobfuscation by code matching. However, layout obfuscation is the easiest, as it does not alter the program logic. 

\subsection{General Code Deobfuscation}

Binary code deobfuscation has been widely studied for the PC platforms (x86)~\cite{guillot2010automatic,likarish2009obfuscated,madou2006loco,yadegari2016automatic,kotov2018towards,david2017formal,gabriel2014deobfuscation,D-OLLVM2017,udupa2005deobfuscation,yadegari2015generic}.
The closest work related to ours might be the one proposed by Gabriel~\cite{gabriel2014deobfuscation}, which attempted to recover O-LLVM obfuscated code on the x86 platform. The work is based on the Miasm~\cite{miasm} framework to reverse the obfuscated function to a control flow graph. Similarly, some work ~\cite{D-OLLVM2017} ~\cite{decompiler_2017}
used similar idea based on different kind disassembler ~\cite{binaryninja, jeb}.
However, they did not tackle the challenges introduced by basic block splitting and instruction optimization, which pose a great challenge for both x86 and ARM platforms. Besides, they did not provide an effective method to identify the obfuscation techniques used in a binary. Finally, context inheritance and sub-function calls are not considered in their work, which is more likely to fail when analyzing large programs. Yadegari \textit{et al.}~\cite{yadegari2015generic} proposed a generic approach to automatically deobfuscate binary code on x86. They use the execution trace extracted from a specific dynamic analysis environment as the input of their system. As a result, their work is not suitable on analyzing shared libraries in Android projects because most shared libraries lack of main functions or entry points~\cite{wiki:Library}.

\section{Approach Overview}
% \section{Native Code Deobfuscation}
% 我们对于ollvm的每一个pass都做了对应的恢复手段，值得注意的是，由于指令替换这个pass是最为简单的混淆手段，在我们的恢复手段中并不会针对他的loop>1的情况进行讨论，instead, our approach only focus on the official suggested setting.

% 这一节我们会介绍我们的方法的具体实现，首先将对反混淆系统的整体框架进行介绍，然后针对ollvm的每个pass，逐一介绍针对该pass的反混淆，各个系统是如何配合的
% In this section, we will elaborate our deobfuscation approach in specific details. Firstly, we will illustrate the overall framework of our O-LLVM deobfuscating system. 

% Second, because instruction substitution and bogus control flow are both semantic levels of instruction obfuscation, while control flow flatten is an overall change to the function control flow. In order to explain our method in detail and make it easy to understand, we will introduce how system works to deobfuscate each pass of O-LLVM separately.

% After that, due to the integrity of our deobufuscation system and the unknownness of actual input, we will illustrate the specific process of the actual situation analysis. 

% \haoyu{TODO: may need rewrite this part.}

% \subsection{System Architecture}

\textbf{Goal.}  We use the term \textbf{deobfuscation} to refer to the process of removing the effects of obfuscation from the native binary, and ideally recover the original code and control flow before obfuscation. For a given APK input, DiANa first extracts the native binary and determines whether it is obfuscated, then analyzes and transforms the code to obtain a functionally equivalent form that is simpler and easier to understand. For non-trivial code, the deobfuscation results is rarely the same as the original code, however, it is close to the original and much easier to understand compared to the obfuscated version.
% we use taint analysis to address the challenge that introduced by instruction optimization. 

\textbf{Key Techniques.}
To address the aforementioned challenges, we rely on taint analysis and enhanced symbolic execution to recover O-LLVM obfuscated code. 
Taint analysis is used to address the challenge introduced by instruction optimization, which performs global feature matching to comprehensively detect all obfuscations introduced by O-LLVM and identify instructions needed to be rewritten by tracking tainted registers.
Symbolic execution is used to reconstruct the control flow ruined by CFF. We first identify basic blocks that maintain original operations based on an ARM-specific basic block classification approach. Then we perform chopped symbolic execution \cite{trabish2018chopped} to address the path explosion challenge. Finally, we use dynamic queue scheduling to maximize the context inheritance and rebuild the original control flow by conducting flow-sensitive symbolic execution.
% 1\ chopped,2\arm based
% To solve the challenge that introduced by basic block splitting, an ARM-based classification approach is proposed. Due to the recover process is flow-sensitive, we use a dynamic exchanging model to make recovery more accurate. 

\begin{figure}[t]
\centerline{\includegraphics[width=0.3\textwidth]{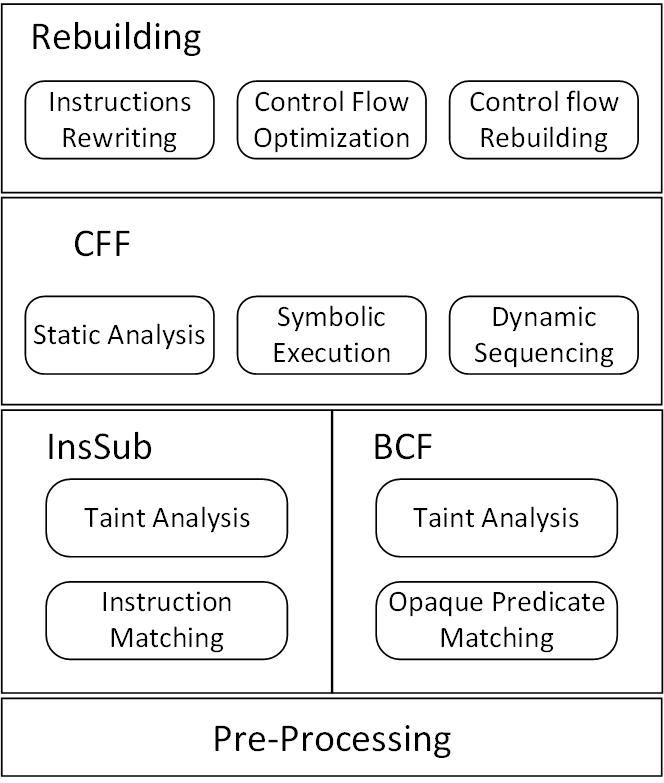}}
%%\vspace{-0.1in}
\caption{Overall architecture of DiANa.}
%%\vspace{-0.2in}
\label{system}
\end{figure}

\textbf{Overall Architecture.}
Fig. \ref{system} illustrates the system architecture of \Name that is mainly composed of four parts. During the pre-processing stage, \Name first determines what kinds of obfuscation techniques are used in the input binary. Then it performs semantic level deobfuscation for InsSub and BCF based on mainly taint analysis. For the binaries protected by CFF, \Name will perform chopped flow-sensitive symbolic execution to rebuild the control flow. Then the recovered control flow will be optimized in consideration of basic block splitting. All the deobfuscation results will be integrated and rewritten to the binary(or control flow graph) in the end.

Next, we will elaborate on the details of \Name and how it works on each obfuscation technique in O-LLVM.

\section{Instruction Substitution Deobfuscation}

The basic idea of InsSub is to replace standard binary operations by functionally equivalent but more complicated sequences of instructions. Note that in order to avoid the interference of constant folding~\cite{wiki:confold}, we set all variable as unknown numbers in our analysis, e.g. a variable waiting for user's input, in order to keep InsSub activated. Table~\ref{Tab:InsSubtable} summarizes 13 kinds of instruction substitution in 5 different categories, which are all the transformations we can find in O-LLVM. Among these transformations, three cases (which are highlighted in the table) are new and not specified in previous work~\cite{junod2015obfuscator}.

\begin{table}[tbp]
    \centering

  \begin{threeparttable}
  \caption{Semantic features used in InsSub.}
  \vspace{-0.in}
  \label{Tab:InsSubtable}

    \begin{tabular}{c|c}
    \toprule
\multicolumn{1}{l}{Obfuscated Expression} & \multicolumn{1}{l}{Origin Expression} \cr
\midrule
{a = b - (-c)} &  \cr
{a = -(-b + (-c))} & \cr
{a = b + r; a += c; a -= r} & \cr
{a = b - r; a += c; a += r} & \multirow{-4}{*}{a = b + c}\cr
\midrule
{ a = b + (-c)} & \cr
 \textbf{\underline{a = -((-b) + c)}} &  \cr
a = b + r; a -= c; a -= r &  \cr
 a = b - r; a -= c; a += r & \multirow{-4}{*}{ a = b - c} \cr
\midrule
a = (b \(\oplus\) ! c) \& b &  \cr
{ \textbf{\underline{a = !(!b \(|\) !c) \& (r \(|\) !r)}}} & \multirow{-2}{*}{a = b \& c} \cr
\midrule
a = (b \& c) \(|\) (b \(\oplus\) c) & a = b \(|\) c \cr
\midrule
{a = (!b \& c) \(|\) (b \& !c)} & \cr
\textbf{\underline{a = (b \(\oplus\) r) \(\oplus\) (c \(\oplus\) r)}} & \multirow{-2}{*}{ a = b \(\oplus\) c} \cr
   \bottomrule
    \end{tabular}
    \end{threeparttable}
\end{table}

% on ARM platform, if the value of a variable was concrete, the InsSub will be optimized by NDK
% \haoyu{CITE HERE}
% \haoyu{Should describe the influence of this NDK optimization. Is is a challenge?}

%  \haoyu{bad presentation here. should discuss to find a better way.} \haoyu{Should describe why.}
% InsSub performs obfuscation on semantic level
\textbf{Challenges.}  
For each kind of operation, there are multiple ways to replace it with functionality equivalent instructions.
To achieve code diversification in the final results, the InsSub obfuscator randomly chooses one way to do the obfuscation. Besides, the obfuscated instructions are often interleaved with normal instructions (unobfuscated ones) at the assembly level, which is hard to separate. As a result, it is not enough by simply searching for specific opcode combinations during InsSub deobfuscation. 
Instead, we introduce static taint analysis to address this challenge.
\textbf{Method.}
We apply taint analysis to determine the combination of obfuscating instructions and locate instructions needed to be rewritten. 
% The analysis process of this subsystem is mainly divided into two steps. First is the opcode combination determining and then is the operands matching. 
We use the following example to illustrate how to use taint analysis to deobfuscate InsSub.

% Taint analysis is used to \Mark{mark to do}. 

%   MOV		R0, R4
  
%   BL		printf
  
%   ADD		R0, R7, R5
  
%   SUB		R6, R0, \#1
  
%   ADD		R0, R0, \#1
  
%   SUB		R7, R0, R5

%   B			loc\_8A0

% Lines 4,6 and 7 form a set of confusion for the addition operation.  
% It works with code simplification transformer to optimized the raw code. We use this code to discuss two different situations.
% \haoyu{mark todo}
% Taint Analyzer consists of two steps: opcode combination determining and operand matching. 

As shown in Fig.~\ref{insObfuscated}(a), assuming line 4 does not exist, the operations from line 3 to line 6 will lead to 
$R7 = R7 + R5 + 1 - R5 = R7 + 1$. After matching the opcode sequence $ADD, ADD, SUB$ and identifying the relationship between these operands, \Name will directly deobfuscate the assembly instruction sequence to its simplest form, $ADD R7, R7, \#1$. Finally, line 5 and line 6 will be replaced by NOP. 
% \lei{modified by Lei}

However, the existence of line 4 indicates that the value of $R6$ depends on $R0$. In this case, \Name will set the modified register $R0$ in line 3 as a tainted register. If there is an instruction operating with the tainted register between or after obfuscated instructions, the involved obfuscated instruction will be retained. As a result, instruction in Line 3 will be retained and instruction in Line 5 and 6 will be set as ``to be patched''. The tainted register will be freed after being modified by the follow-up instruction that is not related to this obfuscating procedure.

\lstset{language=c, label=inscon,
    basicstyle=\scriptsize\ttfamily,captionpos=b,escapeinside=``}
% {\begin{center}
% \vspace{-0.45in}
% \begin{minipage}[t]{1.03\linewidth}
% \begin{lstlisting}
% 	  case addr1:
% 		`\textbf{ToBoldThisLine((val\&0xF)!=(val|0xF), 1); }`
% 		....
% 	...}
%   }else{...}
% }
% \end{lstlisting}
% \end{minipage}
% \end{center}
% }

\begin{figure}
\centering
{\begin{minipage}{0.475\columnwidth}
\begin{lstlisting}[language={[x86masm]Assembler},basicstyle=\footnotesize]
---------------------
1 MOV	R0, R4
2 BL	printf
3 ADD	R0, R7, R5
4 SUB	R6, R0, #1
5 ADD	R0, R0, #1
6 SUB	R7, R0, R5
7 B	loc_8A0
---------------------
(a) Obfuscated
\end{lstlisting}
\end{minipage}}
{\begin{minipage}{0.475\columnwidth}
\begin{lstlisting}[language={[x86masm]Assembler},basicstyle=\footnotesize]
---------------------
MOV	R0, R4
BL	printf
ADD	R0, R7, R5
SUB	R6, R0, #1
`\textit{\textbf{ADD}}`    `\textit{\textbf{R7, R7, \#1}}`
`\textit{\textbf{NOP}}`
B	loc_8A0
---------------------
(b) Deobfuscated
\end{lstlisting}
\end{minipage}}
\vspace{-0.1in}
\caption{An example of InsSub deobfuscation.}
\vspace{-0.2in}
\label{insObfuscated}
\end{figure}

% \begin{figure}
% \centering
% \includegraphics[width=0.6\columnwidth]{approach_pic/BCF-new.png}
% \includegraphics[width=0.3\columnwidth]{approach_pic/BCF-re-new.png}
% \label{opaque}
% \caption{An Example of BCF Deobfuscation.}
% \end{figure}

% \begin{minipage}{0.5\columnwidth}
% \centering
% \begin{lstlisting}[language={[x86masm]Assembler},numbers=left,basicstyle=\footnotesize]
% obfuscated:		simplified:
% ---------------------- --------------------
% MOV	R0, R4		MOV	R0, R4
% BL	printf		BL	printf
% ADD	R0, R7, R5	ADD	R0, R7, R5
% SUB	R6, R0, #1	SUB	R6, R0, #1
% ADD	R0, R0, #1	ADD	R7, R7, #1
% SUB	R7, R0, R5	NOP
% B	loc_8A0		B	loc_8A0
% ---------------------- --------------------
% \end{lstlisting}
% \end{minipage}

% \subsubsection{Rewrite Binary}
% \haoyu{mark to do}
The deobfuscation result of InsSub is shown in Fig.~\ref{insObfuscated}(b). The content and address of the instructions that need to be modified will be saved temporarily and wait until other deobfuscating work finished, before rewritten to the binary.

\section{Bogus Control Flow Deobfuscation}

% \begin{figure*}[]
% \centerline{\includegraphics[width=0.3\textwidth]{pictures/BCF.png}}
% \caption{An example of Instruction Analyzer}
% % \vspace{-0.15in}
% \label{fig 2}
% \end{figure*}
% \includegraphics[scale=0.5]{BCF.png}
% \newline
BCF obfuscation introduces \emph{opaque predicates}\cite{wiki:opaque}
%\footnote{An opaque predicate~\cite{Opaque} is a predicate \(-\) an expression that evaluates to either ``true'' or ``false'' \(-\) for which the outcome is known by the programmer a priori, but which, for a variety of reasons, still needs to be evaluated at run time.} 
into the functions. 
The opaque predicate expression used in O-LLVM is \textbf{y\(<\)10 or (x*(x-1) mod 2)==0}. The second half of the expression is an odd number multiplying an even number modulo 2, which must equal 0. Hence, the whole expression will always be true, and the obfuscator introduces a dead branch to the code. Note that the branch will not be optimized by compilers due to the uncertainty of \textbf{x}. 
% \haoyu{no optimization}
% A dead branch consist of part of the content of the basic blocks in a real branch.  This is important for the deobfuscation of three pass mixed case, which will be illustrated in the Control Flow Flatten section.

\textbf{Challenges.} 
Although BCF obfuscation has \emph{modified} the control flow (by introducing dead branches), it is still instruction-level obfuscation. As a result, the challenge we encountered is similar to InsSub. However, BCF is significantly more complicated because the obfuscated instructions are often across basic blocks. To address this challenge, we also rely on taint analysis to perform global opaque predicate matching in order to remove dead branches efficiently.
% How to globally match opaque predicate operations has become a challenge for our deobfuscation work. 
% We still rely on taint analysis to address this challenge.
% \haoyu{mark to do. how and why?}
% \Mark{I already said in InsSub}

% \begin{figure}[htbp]
% \centerline{\includegraphics[width=0.4\textwidth]{approach_pic/BCF-new.png}}
% \caption{BCF.}
% \vspace{-0.1in}
% \label{fig 4}
% \end{figure}
\begin{figure}[tbp]
\includegraphics[width=0.9\columnwidth]{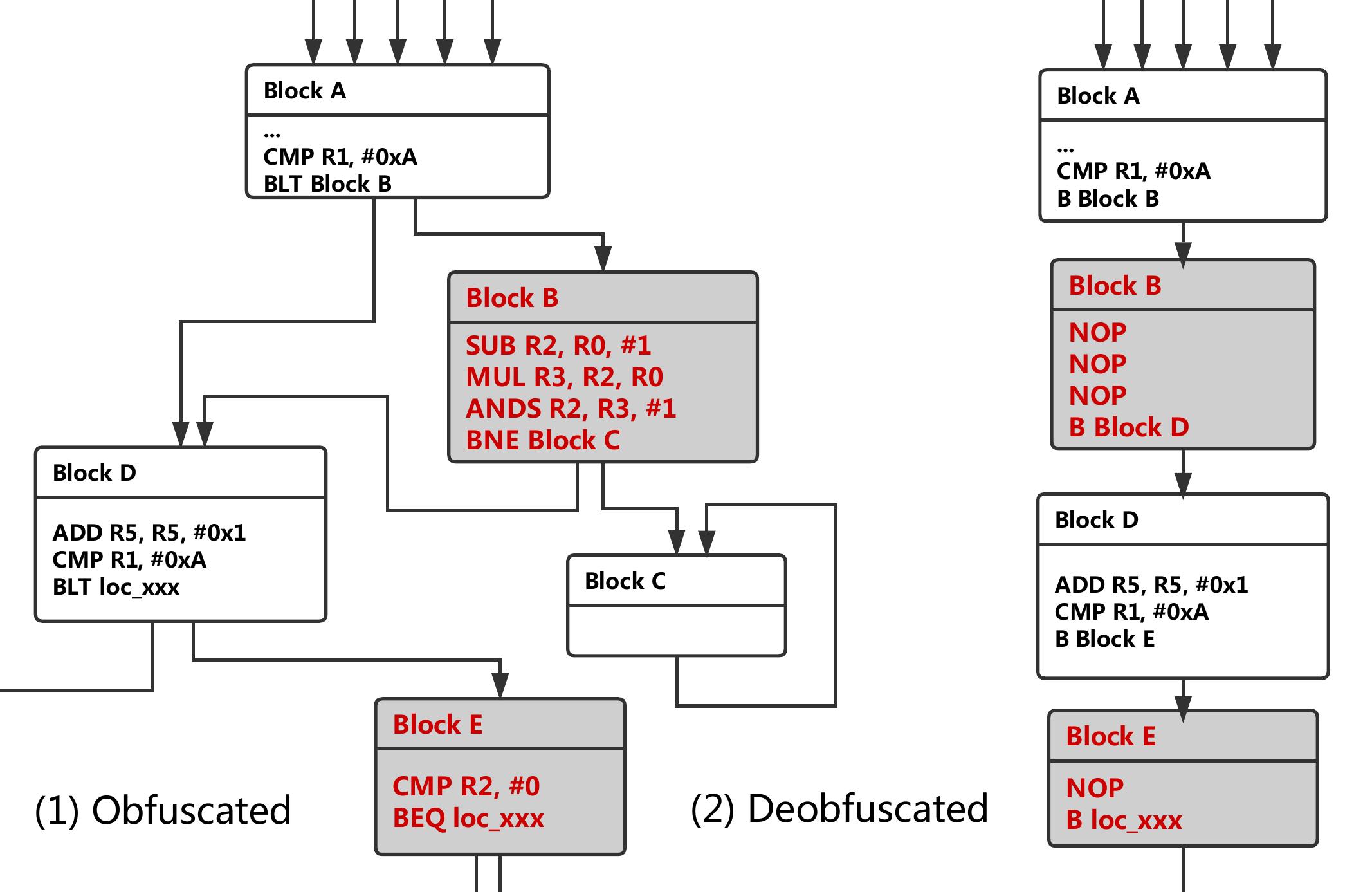}
% \includegraphics[width=0.3\columnwidth]{approach_pic/BCF-after.png}
%%\vspace{-0.15in}
\caption{An example of BCF deobfuscation.}
\label{opaque}
\vspace{-0.1in}
\end{figure}

%  \haoyu{modify this figure.}
% \subsubsection{Determine Features}

% The principle of BCF is a pseudo-addition of control flow on the basis of instructions\haoyu{mark to do}. 

\textbf{Method.}
We rely on assembly features to detect whether a binary is obfuscated by BCF. 
For the given opaque predicate, we have summarized multiple assembly features. Note that InsSub may change the assembly features of BCF (e.g. sub operation may be obfuscated), thus we also manually summarized BCF features when InsSub is involved.

% Feature Extractor first extracts the initial control flow, and check the existence of assembly features of the opaque predicate operation. Once the feature was found, the input binary will be considered as obfuscated by BCF. 
% The initial control flow and all block contents \haoyu{mark to do} will be passed to Taint Analyzer.
% 事实上虚假控制流的原理也是在指令的基础上对控制流进行虚假的添加
% The initial control flow will first be extracted. Then, once the expression of the opaque predicate is determined, Feature Extractor is used to locate all basic blocks with opaque predicate operation \textbf{(x*(x-1) mod 2)==0}. It will save all addresses of these blocks in a list, and hooks all the registers which the opaque predicate changed the value of, to facilitate the follow-up taint analysis. 

% \subsubsection{Remove Bogus Branch}
% \subsubsection{Taint Analysis}
BCF introduces two branches to the original block using `CMP Rx, 0xA' (y$<$10). One successor block containing the `SUB MUL AND(S)' combination (maybe different implementations of the same expression `(x*(x-1) mod 2)' is used to direct the control flow back to the origin block. Another branch of this successor is always headed to a dead branch.
All basic blocks that take opaque predicates as jump conditions are designated as \textit{predicate blocks}. We use Fig. \ref{opaque} as an example to show how \Name works on this obfuscating pass. 
% \haoyu{mark to do. why use taint analysis and how.}

As shown in Fig. \ref{opaque} (1), the `SUB MUL ANDS' sequence first appears in \texttt{Block B}. When the opaque predicate operation is detected, taint analysis first sets \texttt{Block B} as a predicate block and sets R2 as a tainted register. Then it will force the conditional jump in the parent block reaching the predicate block \texttt{Block B}. After that, because only one condition jump of the predicate block can actually be accessed, taint analysis will head to \texttt{Block D} and label jump instruction at the end of \texttt{Block B} as to be modified. 
% The address of aimed alive block will be saved too to determine the actual modified instruction content.

Due to the optimization on ARM instructions during compilation, sometimes BCF obfuscation can occur as shown in \texttt{Block E}. If a conditional jump occurs and the constraint is a tainted register compared to zero (there are some cases that compare with 1), \Name will automatically set it as a predicate block. In this case, because register R2 is tainted, when system traverses the child blocks of \texttt{Block D}, the block \texttt{Block E} will be set as the predicate block too. The tainted register will be freed when modified by normal instructions.

% In subsequent analysis, if a conditional jump occurs and the constraint is a tainted register compared to zero (there are some cases that compare with one), Taint Analyzer will automatically set these blocks as predicate blocks. 

% Only the alive path will be explored in the follow-up process.
% Code simplification transformer will set instruction in Line 6 as the one that needs to be patched.
% The 'SUB MUL AND(S)' combination appears in the second block. As a result register R2 will be set as tainted register and No.2 and No.4 will be set as predicate blocks. Then it will first follow the conditional jump of 'CMP r1, \#A' to jump to the predicate block.

% In order to prevent instructions that affect original control flow from being destroyed, Taint Analyzer will let control flow pass through instead of avoiding predicate blocks. 

% \subsubsection{Rewrite Binary} 

After taint analysis, all deobfuscated information about instructions that need to be modified will be passed to the rebuilding process. When the binary is not obfuscated with CFF at the same time, the deobfuscated result (as shown in Fig. \ref{opaque} (2)) will be rewritten to the binary. 
% A simple deobfuscated result of BCF is shown in the right side of Fig. \ref{opaque}\haoyu{todo}. 
% After removing bogus branches and dead blocks, the original operation of this area is completely preserved.
% \begin{figure}[htbp]
% \centerline{\includegraphics[width=0.4\textwidth]{approach_pic/BCF-re-new.png}}
% \caption{BCF.}
% \vspace{-0.1in}
% \label{fig 4}
% \end{figure}

\section{Control Flow Flattening Deobfuscation}

% The CFF obfuscation removes all easily identifiable conditional (e.g., \texttt{IF-ELSE}) and looping structures (e.g., \texttt{WHILE,FOR}), and reconstructs control flow through the use of a large \texttt{SWITCH} construct, which is responsible to route the code control flow through the proper basic blocks depending on a routing variable~\cite{junod2015obfuscator}.
% The dispatch process \haoyu{todo, rewrite}. 

The CFF obfuscation rebuilds the control flow to a \texttt{SWITCH} construct. To deobfuscate the control flow, we need to identify basic blocks that maintain original operations and rebuild the control flow. Symbolic execution is a means of using symbolic input to analyze a program to determine what inputs cause each part of a program to  execute~\cite{wiki:symbexe}. Since the analysis of CFF obfuscated code is flow sensitive, and symbolic execution has been proved as an effective program analysis technique that can systematically explore multiple program paths~\cite{cadar2011symbolic,anand2007jpf,cha2012unleashing}, we combine automatic static analysis and symbolic execution to deobfuscate the binary code. However, we still need to address the following \textbf{Challenges}:

% Traditional static analysis is hard to explore the branches completely, while dynamic analysis typically are too complex to cover all the branches. \Mark{}
% Symbolic execution is a means of using symbolic input to analyze a program to determine what inputs cause each part of a program to execute~\cite{wiki:symbexe}. It has been proved as an effective program analysis technique that can systematically explore multiple program paths~\cite{cadar2011symbolic,anand2007jpf,cha2012unleashing}.
% \haoyu{cite 2-3 refs here}. 
% \haoyu{one sentence to introduce symexe here.}
% 由于ARM平台本身的优化特性，会对基本代码块

% \haoyu{why they are challenges? should describe more clear here.}\Mark{I wrote in each points}

% As a result, 

%%\vspace{-0.05in}
\begin{enumerate}
\item Due to the instruction optimizations introduced by the ARM compiler and obfuscator, original blocks are often split and reused. It is a challenge to identify the blocks that contain original operations from a super complicated control flow.
% How to identify basic blocks contain original operation from a super complicated control flow? 
% Identifying basic blocks contain original operation from a super complicated control flow. 

\item  As the original control flow has been destroyed by CFF, most basic blocks use direct jumps to reach successors after obfuscation. Before rebuilding the original control flow, we need to know what kind of blocks have multiple successors in the original flow. 

% How to determine the number of original successors of a obfuscated basic block is 
% Preforming context-aware symbolic execution. 

\item The recovery work on CFF-obfuscated code is flow-sensitive (e.g. the switch structure relies on the routing variable, which affects the control flow), thus we need to know how to maximize context inheritance and avoid path explosion.
% Due to the dependency of the switch structure on the routing variable, context information becomes very important in the recovery process. 
% flow sensitive 

\end{enumerate}
%%\vspace{-0.05in}
% We should maximize the context inheritance and avoid the problem of path explosion.
% \Mark{rewrite}

\Name first determines whether the code is obfuscated by CFF during preprocessing, and then conduct static analysis to identify basic blocks that contain original operations (\texttt{OO Blocks} for short). After that, it performs flow-sensitive symbolic execution on these blocks to reconstruct the original control flow. After optimization, the final deobfuscation result will be written to the output.

% we first use static analysis to find out basic blocks that contain original operations. Then symbolic execution is used to recover jumps between these basic blocks.

% XXX Then XXX...\haoyu{summarize the CFF Deobfuscation approach here.}

\subsection{Feature Extraction}
% The first step of our deobfuscation system is to make sure whether a function is obfuscated by CFF, which will be achieved by Feature Extractor. 
Since the flow between basic blocks is directed by a big switch construct, which is led by a routing variable, we use this switch construct and routing variable as two fingerprints to identify CFF-based obfuscation. 
% Based on manually analysis, we found that t
% \texttt{IF-ELSE} 
% After determining CFF is used to obfuscate code, Feature Extractor will extract 

The switch construct is actually a nested set of conditional jumps. As shown in Fig. \ref{CFF Structure}, each block ending with a conditional jump mainly consists of three instructions. We call these blocks \textbf{Dispatchers}. Also, the second operand of the compare instruction, which is different from other compare operations, is determined by both the value of the ARM \textbf{R15} register (Program Counter Register) and an immediate number \(\alpha\). \Name calculates the value of routing variable as follows:
\begin{equation}
V_{routing}=\alpha+ ins.address + 8\label{con:routine}
\end{equation}

% \textbf{routing variable =} \textbf{\(\alpha\)} \textbf{+ instruction.address + 8}

% \footnote{In Thumb mode, the value of PC Register equals current instruction address add 4.}
Here, \texttt{ins.address} is the address of the current instruction, which will be added by 8 (add 4 in Thumb mode), and becomes the value of Register R15 in the ARM mode.
All values of \(V_{routing}\) are saved into a dispatcher dictionary. 
%  which we called \textbf{\textit{KEY DICT}}

\subsection{Identification the OO Blocks}

\begin{figure}[tbp]
\centerline{\includegraphics[width=0.5\textwidth]{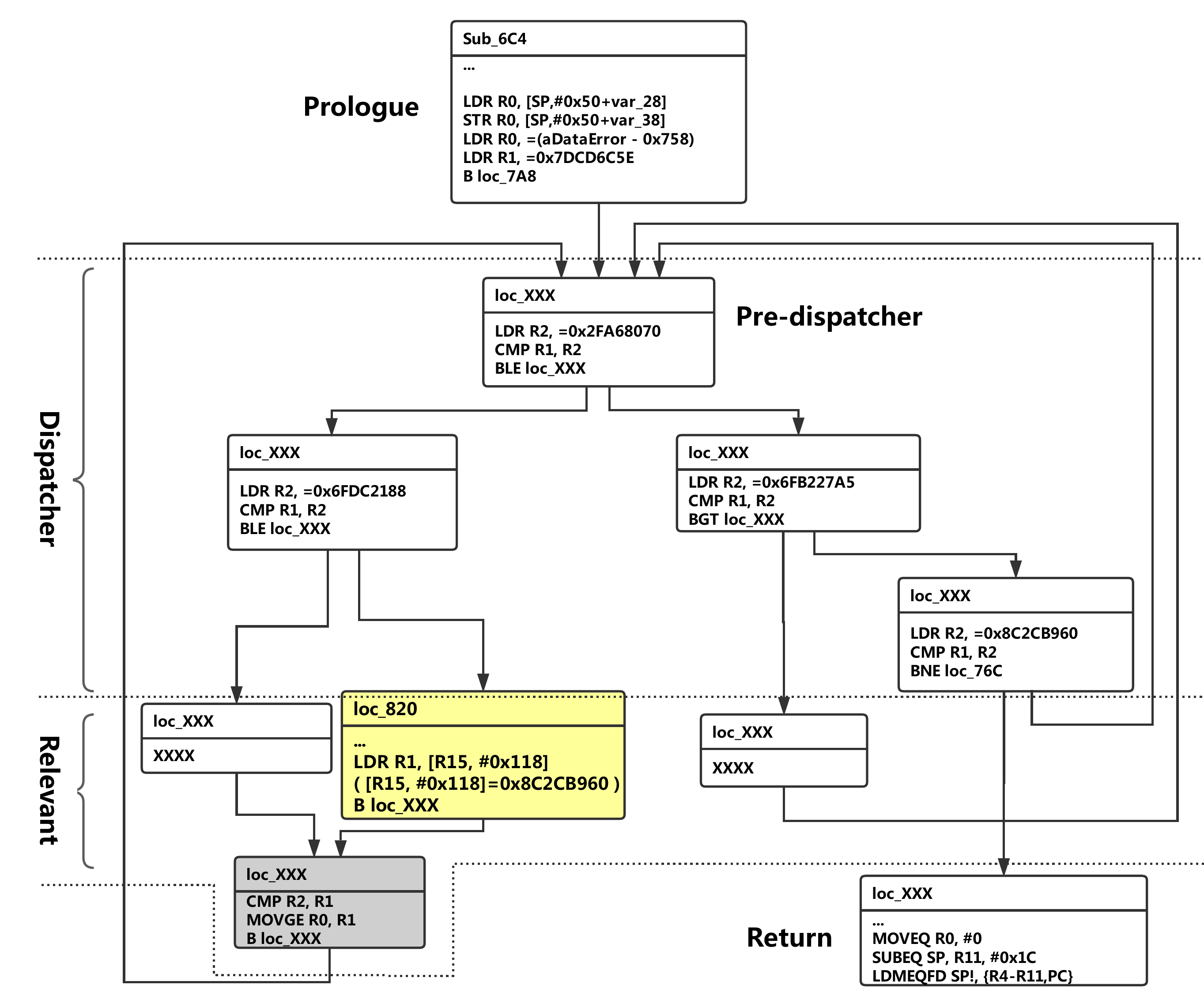}}
\vspace{-0.1in}
\caption{An example of a CFF obfuscated function.}
\vspace{-0.2in}
\label{CFF Structure}
\end{figure}

Though CFF ruined the original control flow, original operations still exist in part of the obfuscated basic blocks. 
To re-construct the control flow, we first need to identify the \texttt{OO Blocks}, i.e., the blocks containing the \textbf{\textit{Original Operations}}.
In reference to previous work on the x86 platform~\cite{gabriel2014deobfuscation}, and combined with the characteristics of CFF on ARM, we divided the obfuscated control flow structure into five categories, as shown in Fig. \ref{CFF Structure}.

\textbf{Prologue} is the entrance block that contains information about almost all the constants in the original function. 

\textbf{Dispatcher} is a conditional-jumping block whose constraint is determined by R15 and an immediate number.
% which contain elements in KEY DICT. 

\textbf{Pre-dispatcher} is the Dispatcher block whose in-degree is larger than two. Note that an obfuscated function may have more than one Pre-dispatchers. 
% It is often same as the return block of the unobfuscated function. 

\textbf{Return} is the basic block whose out-degree is 0. Note that an obfuscated function may have more than one Return blocks.
% The search for Relevant blocks is the most important thing in the entire deobfuscating process, and it decides the correctness of the following-up work.

\textbf{Relevant} is the basic block that maintains operations of the original function. 
\Name does a reverse traversal from the Pre-dispatcher until it encounters a Dispatcher block. All basic blocks in this path between the Pre-dispatcher and Dispatcher are Relevant blocks because these blocks are all dispatched by the switch structure.

The original operations exist in Prologue, Return and Relevant Blocks, which forms the \texttt{OO Blocks}. 
Due to the introduced optimizations,  Relevant blocks may be partly merged, as the bottom gray block shown in Fig. \ref{CFF Structure}. In this case, the block will be duplicated and added to its two parent nodes.
% Taint Analyzer does a reverse traversal from the Pre-dispatcher until it encounters element in Dispatchers. In these resulting paths, all basic blocks between the Pre-dispatcher and Dispatcher are Relevants. Due to the optimization on ARM, some Relevants may partly merged, as the gray block in Figure.\ref{CFF Structure}. In this case, gray block will be duplicated and added to its two parent nodes.

%  a Relevant block has multiple branches or not
\textbf{Determine the number of original successors.}
Before the recovering process, we need to determine whether an \texttt{OO Block} has multiple successors when it was in the original control flow. 
% Note that Prologue block has one or two branches, ASSUME. \haoyu{todo here} Thus, we only need to analyze the Relevant blocks. 
The goal of this step is to prepare for subsequent analysis. Symbol execution will be performed twice to explore each possible path if a basic block has two successors.
% \haoyu{rewrite}

% Once a Relevant block has instruction which assigns a register, and the second operand of this instruction is dependent on R15 register, we calculate the result use Formula.\ref{con:routine}. If the calculation result exists in the range of dispatcher dictionary

Different from previous work on the x86 platform~\cite{gabriel2014deobfuscation}, which just used the opcode `CMOV' as the feature to detect the existence of multiple branches, and due to the simplicity of the ARM instruction set, we use elements in the previously mentioned dispatcher dictionary as the main indicator. If the second operand of an assignment operation exists in the dispatcher dictionary, the assigned register will be set as ``tainted register''. If there is a tainted-register-involved conditional move instruction on the path of this block back to Pre-dispatcher, we consider that this block has multiple successors in the original control flow. As previously mentioned, the tainted register will be freed when modified by a normal instruction. As shown in Fig. \ref{CFF Structure}, the yellow block (loc\_820) is an example that has two branches in the original function. 

% In this block, there is an instruction 'LDR R1, [R15, \#0x118]' and the calculation result of \textbf{formula 1} exists in dispatcher dictionary. As a result, R1 will be set as a taint register. Also there is a conditional move instruction 'MOVNE R1, R9' on the path the flow go back to Pre-dispatcher, which involves the taint register R1. According to our theory, this highlight basic block will be judged having two branches.

\subsection{Flow-sensitive Symbolic Execution}

Symbolic execution can be used to systematically explore multiple program paths\cite{trabish2018chopped}. 
However, due to unknown path constraints of the obfuscated program, directly applying symbol execution to explore all possible branches will cause path explosion.
% if the symbol execution is directly performed, it is easy to cause the path explosion problem. 

\subsubsection{Chopped Symbolic Execution}

\mbox{} \\ Inspired by the theory of chopped symbolic execution by David \textit{et.al}~\cite{trabish2018chopped}, which mainly targeted the exploration to paths of importance, we decide to separate the process of symbolic execution and set each \texttt{OO Block} as an object to be analyzed. For each basic block that needs to be analyzed, if it is previously determined to have two successors, symbolic execution will be performed twice to find out each of its successors separately. Otherwise the execution will be performed only once.
% 对于每个需要分析的基本块，我们会对其每个后继分别进行查找。如果一个基本块之前被判定存在两个后继，则对他的符号执行将会进行两次。
We use Angr~\cite{shoshitaishvili2016state} as our symbolic execution engine to perform symbolic execution on each \texttt{OO Block} in turn. 

% Here, }, a XXXX tool\haoyu{todo here} .

% In fact, Dispatcher and Pre-dispatcher blocks are added by the obfuscator to build the switch construction. Basic blocks which are useful for our recovery work exist in Prologue, Relevants and Return. We set each useful basic block as analysis object and perform symbolic execution on them in turn. 

% 子标题要改
% 一段保留，2过程
% \item \textbf{The order of symbolic execution.}\newline
% In fact, Dispatcher and Pre-dispatcher blocks are added by the obfuscation tool to build the switch construction. Basic blocks which are useful for our recovery work exist in Prologue, Relevants and Return. 

\subsubsection{Execution Process}

\mbox{} \\ \Name first pushes Relevant blocks into a block sequence. Initially, these blocks are out of order. The Prologue block is the first one to be analyzed. Note that in \Name, we always assume that Prologue has two succeeding branches. As a result, \Name performs symbolic execution twice on Prologue to explore both branches. We use the break point function in Angr to interrupt the execution before conditional \texttt{move} instructions. By changing the execution state, it will let control flow enter both True and False branches.
% \haoyu{explain why} 
% add \texttt{BP\_BEFORE} type breakpoint to the \texttt{statement} attribute. 
%  and execute the action function we set. When the symbolic execution is interrupted
% \texttt{claripy.BVV(1,1)}
% \texttt{claripy.BVV(0,1)}
% Here \texttt{claripy.BVV(v, len)} creates a \textbf{\textit{len-bit}} vector with the value of \textbf{\textit{v}}. 

When the execution stops before the break point, our system will set the conditional flag in Angr's Intermediate Representation to symbol 0 and 1, respectively, for the purpose of exploring its both following branches. For blocks with a single branch, \Name does not modify the state and analyzes them directly. The symbolic execution will keep stepping on until reaching an element in the block sequence or Return. 

% As a result, before the execution, our system will firstly traverse each instruction in this basic block and its son blocks.
\subsubsection{Cross-Function Protection}

\mbox{} \\ During symbolic execution, it is possible to enter another function's space through the function call. In ARM instructions, calling functions are done via \texttt{`BL'} or \texttt{`BLX'}. \Name hooks these two call instructions. When symbolic execution encounters hooked addresses, it will automatically skip the call instruction and continue the analysis in the space of the current function.

% todo
{
\begin{algorithm}
\caption{Dynamic Queue Scheduling Algorithm.}
\LinesNumbered %要求显示行号
\KwIn{Prologue \textit{P}, Relevants \textit{R}, Return \textit{Q}}%输入参数
\KwOut{Recovered control flow \textit{F}}%输出

    BlocksQueue[0] = \textit{P}\;
    BlocksQueue[1: ] = \textit{R} + \textit{Q}\;
    % OO\_Blocks = Prologue + Relevants + Return\;%\;用于换行
    Swap\_Pointer $\Leftarrow$  BlocksQueue[1]\;
    
    % successor1, CurrentState1 = SymbExec (Prologue, flag=true)\;
    % SavedState[successor1] = save (CurrentState1)\;
    % F[Prologue] = F[Prologue] + successor1\;
    
    % successor2, CurrentState2 = SymbExec (Prologue, flag=false)\;
    % SavedState[successor2] = save (CurrentState2)\;
    % F[Prologue] = F[Prologue] + successor2\;

    % Execution\_Pointer = BlocksQueue[0].address\;
    % swap(successor1, 

\For{i \textbf{in} range(0: len(BlocksQueue))}{
    BB = BlocksQueue[i]\;
    \eIf{BB has saved\_state}{
        RecoverState(SavedState)\;}
        {RecoverState(BlankState)\;}
    
    successor\_num = find\_branch(BB)\;

        \eIf{successor\_num == 1}{
            successor, CurrentState = SymbExec (BB, flag=None)\;
            SavedState[successor] = save (CurrentState)\;
            \textit{F}[BB].successor += successor\;
		\If{successor.address > Swap\_Pointer}
			{
			    swap(successor, Swap\_Pointer)\;
			    Swap\_Pointer++\;
			}
        }{
            successor1, CurrentState1 = SymbExec (BB, flag=true)\;
            SavedState[successor1] = save (CurrentState1)\;
            \textit{F}[BB].successor += successor1\;
            \If{successor1.address > Swap\_Pointer}
			{
			    swap(successor1, Swap\_Pointer)\;
			    Swap\_Pointer++\;
			}
            
            successor2, CurrentState2 = SymbExec (BB, flag=false)\;
            SavedState[successor2] = save (CurrentState2)\;
            \textit{F}[BB].successor += successor2\;
            \If{successor2.address > Swap\_Pointer}
			{
			    swap(successor2, Swap\_Pointer)\;
			    Swap\_Pointer++\;
			}
        }
    }

\Return{\textit{F}}
\end{algorithm}
}

\begin{figure}[t]
\centerline{\includegraphics[width=0.5\textwidth]{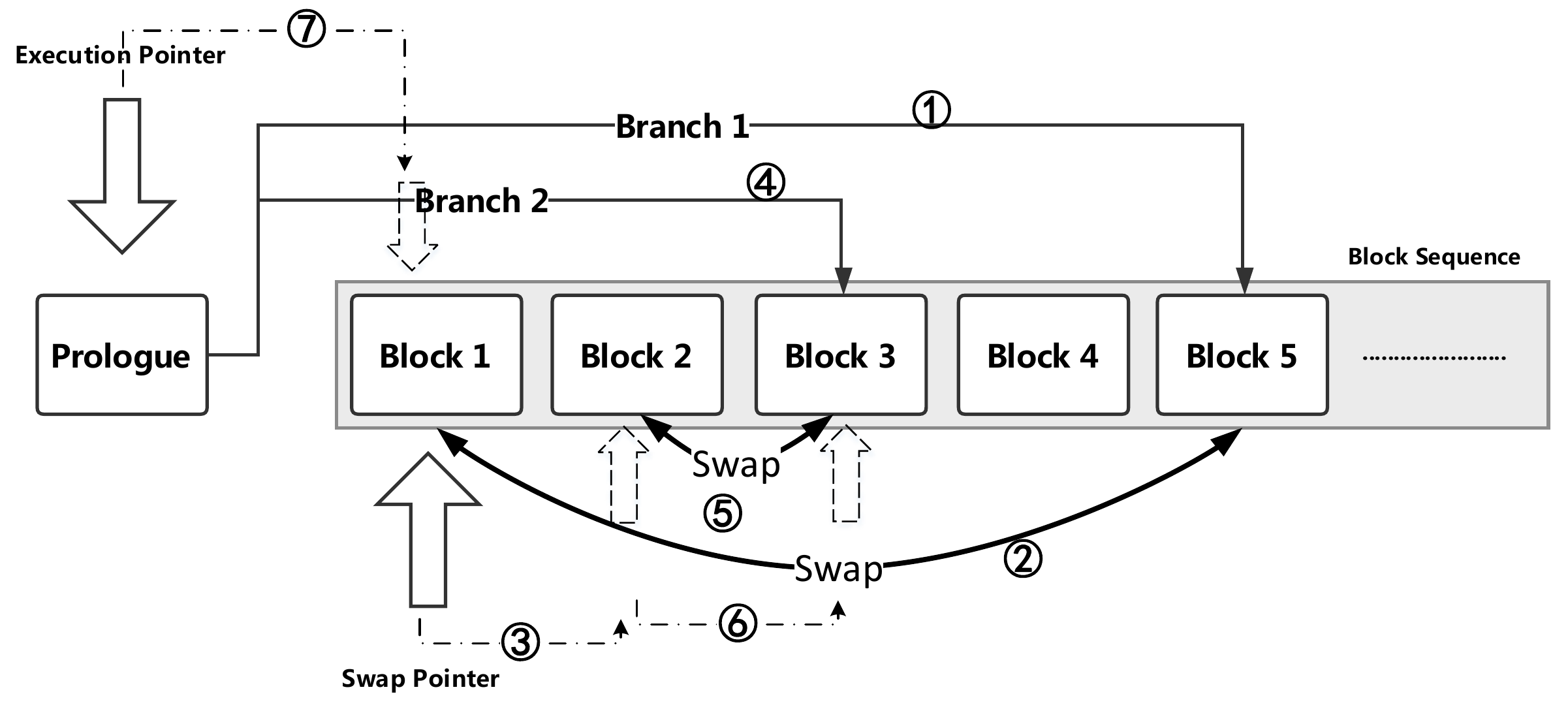}}
\vspace{-0.1in}
\caption{An example of dynamic queue scheduling.}
\vspace{-0.3in}
\label{fig:Symbexe}
\end{figure}

\subsubsection{Flow-Sensitive Execution}

\mbox{} \\ Since there are dependencies between basic blocks in a function, the deobfuscation process is flow-sensitive. If the symbolic execution starting from each basic block uses a blank symbol state, it must cause basic block mismatching due to the loss of necessary context. 
% \haoyu{explain the challenge here} 
As a result, while performing symbolic execution on basic blocks, the deobfuscation system will dynamically save current execution state and adjust the position of the elements in block sequence to ensure the state inheritance. 
% they are \texttt{Execution Pointer} and \texttt{Swap Pointer}.

Two pointers are used in the process of symbolic execution: the \textit{Execution Pointer} pointing to the current analyzing block, and the \textit{Swap Pointer} pointing to the front position that never being swapped. Initially, the Execution Pointer points to the Prologue and the Swap Pointer points to the first block in the block sequence. The returned successor block will be swapped to the position that the Swap Pointer points to. At the same time, the current symbolic state will be saved. When the analysis proceeds to this successor, the saved state will be restored. If the returned successor block is before the Swap Pointer, there will be no exchange between basic blocks. But the symbolic state of this successor will be updated. After getting all successor block(s) of the current analyzing block, the Execution Pointer will move to the next block. In special cases when the two pointers coincide, \Name will move the Swap Pointer forward one space.

The purpose of dynamically exchanging basic blocks in the sequence is to ensure that most of the basic blocks inherit the state from the previous analysis before execution, while not starting the analysis of a basic block from a blank state. 

Algorithm 1 presents the top-level algorithm of our dynamic basic blocks scheduling approach. We also use Fig.~\ref{fig:Symbexe} as an example to show how this algorithm work. \textcircled{1} After analyzing the True branch, the state before executing Block 5 is saved. \textcircled{2} Then Block 5 is exchanged to the first place of sequence and \textcircled{3} the Swap Pointer is moved to the next. \textcircled{4} Following the analysis of the False branch, \textcircled{5} the successor Block 3 is swapped with the second block in sequence and \textcircled{6} the Swap Pointer is moved to the third position. At the same time the state reaching Block 3 will be saved. After that, \textcircled{7} the Execution Pointer points to the first block (Block 5) in the sequence and the previously saved state of Block 5 will be restored. The analysis of other blocks will continue like this successively.

Performing symbolic execution in this way not only guarantees the maximum state inheritance of each basic block, but also avoids path explosion and ensures the order of the entire analysis process.

\subsection{Control Flow Reconstruction}
When using CFF to obfuscate the native code, there is a certain degree of basic block splitting and merging due to ARM optimizations, no matter the basic block splitting pass is activated or not. As a result, it is necessary to optimize the recovered control flow. We apply two optimization here.
% The optimization is mainly concentrated in two parts, as shown in Fig. \ref{fig:mergeRule}.

\begin{figure}[htbp]
%\vspace{-0.2in}
\centering
\includegraphics[width=\columnwidth]{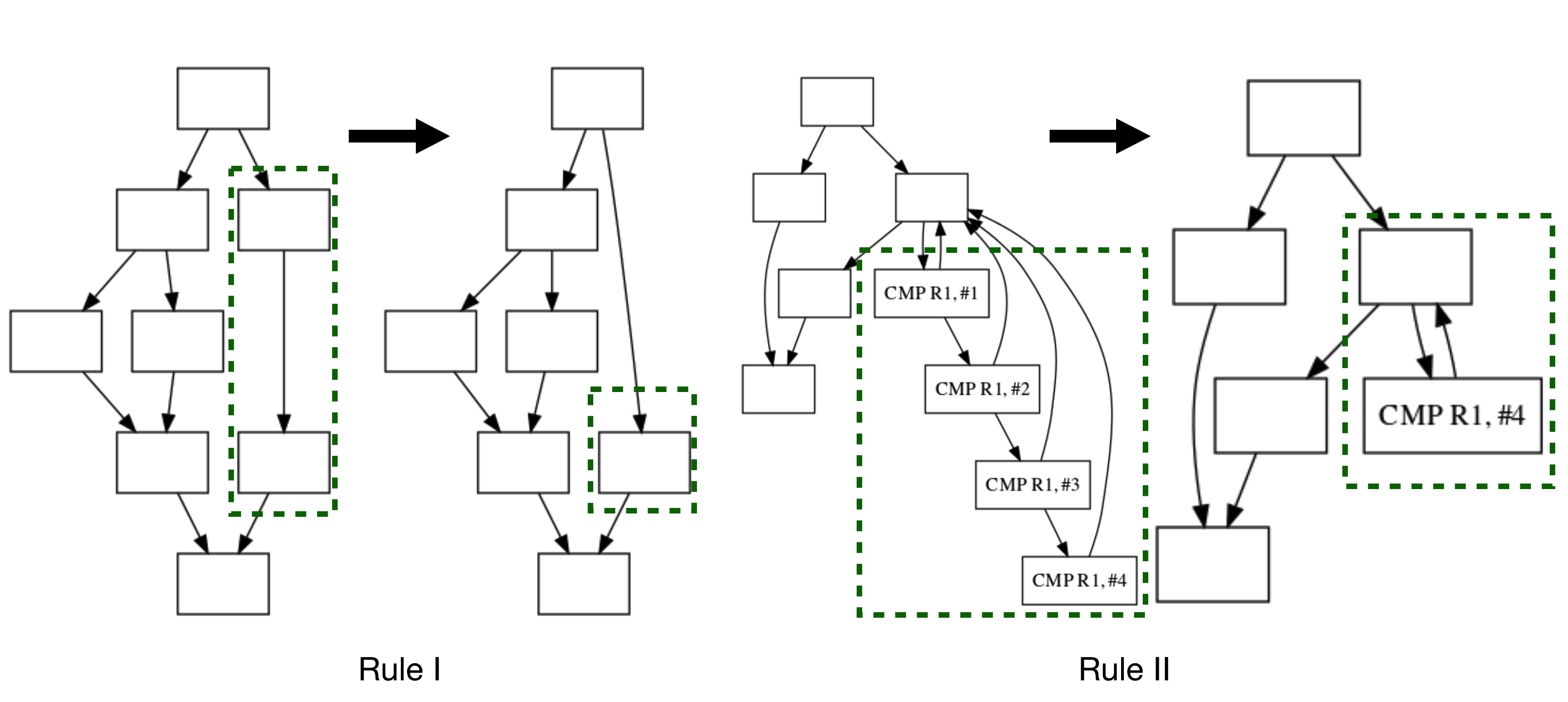}
\vspace{-0.2in}
\caption{Control flow merging rules.}
\label{fig:mergeRule}
\vspace{-0.2in}
\end{figure}

\uppercase\expandafter{\textbf{Rule \romannumeral1}}. As shown in Fig. \ref{fig:mergeRule} (1), for two nodes connected by a direct jump, if the in-degree of the parent node is not greater than one, \Name merges these two nodes.

\uppercase\expandafter{\textbf{Rule \romannumeral2}}. As shown in Fig. \ref{fig:mergeRule} (2), for multiple connected nodes, if their out-degree are all equal to two and the other branch of them points to a same node, once the contents of each node are alike (e.g. comparing operation with consecutive integers), \Name will optimize this structure as a loop.
% for multiple nodes with two branches, 
% if these nodes are connected in series by one branch, and their other branches point to the same node, once the opcode sequences are identified, 

Note that when BCF and CFF are used together (with or without InsSub), the introduction of opaque predicates does not affect the recovery of original control flow. Because these blocks (previous mentioned blocks in dead branch) are always unreachable, they will be removed in the final output.

\section{Implementation and Evaluation}

% \subsection{Implementation}

% Our experiment is carry out on a machine with 2.9 GHz Intel Core i7 processor and 16 GB memory running macOS Version 10.13.6. When we did our research, the latest version of O-LLVM was 4.0, and all of our work was based on this version's obfuscator. It is noticeable that our approach is suitable for code that obfuscated by any OLLVM versions. 
Our experiment is based on O-LLVM 4.0, while our approach can be easily extened to code obfuscated by any OLLVM versions. We use Android NDK V16.1 to build shared libraries. The symbolic engine of DiANa is implemented based on Angr 7.8.2, which is inspired by existing work on x86 platform~\cite{CFFRef, CFFRef2}.
The function addresses are exported by IDA Pro 7.0.

\subsection{\textbf{Dataset}}

To evaluate the effectiveness of DiAna, we first need to build a reliable open source dataset with high coverage. We have applied our system to the following three different datasets, respectively. 

\textbf{C/C++ Obfuscation Benchmark.}
We first evaluate our system on a widely used C/C++ obfuscation benchmark~\cite{cbenchmarks}. This dataset was created to evaluate different obfuscation algorithms~\cite{banescu2016code}. It includes common basic algorithms, hash functions, and small programs (mainly containing combinations of \texttt{IF}, \texttt{WHILE}, and \texttt{FOR} structures). We eliminated the benchmark programs that contain only one basic block, because it is hard to apply obfuscation techniques on these tiny programs. At last, we obtain 94 benchmark programs in total to evaluate the effectiveness of DiANa.

\textbf{Open-Source Android Native Code.} 
To evaluate our system on real-world Android apps, we have crawled 100 open source Android projects from F-Droid~\cite{F-Droid}. Among them, 24 projects use native code. Note that Android apps could use native code in two ways: reusing the existing native libraries by embedding the $*.so$ in the project or implementing their own C/C++ code and then compiling it within the project. Finally, we have identified 5 real-world Android projects that contain open source C/C++ code, which can be used to evaluate our approach. 

\textbf{Real-world Andrioid Root Exploit.} 
We further apply \Name to Android Root Exploits (aka REs), a binary tool used to obtain the root privilege by exploiting privilege escalation vulnerabilities. In this case, we could evaluate the effectiveness of \Name on deobfuscating real threats in the wild.

% As many malware may hide the malicious payload in the native code to evade detection, we found a real-world malware case that exploit the towelroot vulnerability to XXXX. 
% \Mark{TODODO DOD OODO DDOO DOD ODO}

%%\vspace{-0.05in}
\subsection{\textbf{Evaluation Metrics}}
We hereby define our evaluation metrics. We deployed \texttt{Euclidean Distance (ED)}, \texttt{Similarity of Control Flow Graph (Sim)}, and  \texttt{Input/Output equivalence (I/O)} to evaluate the effectiveness and correctness of \Name.  Because InsSub works at the assembly instruction level and does not alter the control flow, we use \texttt{ED} to measure the deobfuscation effectiveness and \texttt{I/O} to ensure the correctness. For the BCF and CFF approaches, we deployed \texttt{Sim} to quantify the deobfuscation results, similar to previous work~\cite{yadegari2015generic}. Most importantly, we also use \texttt{I/O} to ensure the deobfuscated function is semantically equivalent to the original one. %Following is the detailed introduction to the three evaluation metrics.

\subsubsection{Euclidean Distance}
\mbox{} \\ 
We use Euclidean Distance as a quantitative indicator of our work on Instruction Substitution. To evaluate InsSub deobfuscation, we first generate a feature vector for each function. Each dimension in the vector represents the frequency of certain opcode. 
% We omit the details of calculating Euclidean Distance in this paper, which could be referred to.
We calculate the Euclidean distance~\cite{wiki:distance} as in Formula \ref{distance}.
\begin{equation}
d(x,y):=sqrt(\sum_{i=1}^{n}(x_i-y_i)^2)\label{distance}
\end{equation}

Here, $x$ and $y$ are the feature vectors of the obfuscated/deobfuscated function and the original one respectively, while $n$ is the total number of types of opcode used in functions $x$ and $y$.

\subsubsection{CFG Similarity Comparison}

\mbox{} \\ 
From the perspective of reverse engineering, control flow analysis is an important procedure. Due to the difficulty of patching (rewriting) the CFF obfuscated native code at the assembly level, CFG analysis is a key technique in static analysis for Android apps. We adopted the algorithm of Hu~\cite{hu2009large} for computing the edit distance between two CFGs, which was suggested as one of the best CFG comparison algorithms in previous empirical study~\cite{chan2014method}. It has also been widely used in previous work \cite{yadegari2015generic}\cite{pawlowski2016probfuscation}\cite{fyrbiak2018hybrid}.

The basic idea of this algorithm is to build a cost matrix that represents the costs of mapping different nodes in two graphs \textbf{\(G_1\)} and \textbf{\(G_2\)}. Then the Hungarian algorithm~\cite{kuhn1955hungarian} is used to find an optimal solution to the assignment problem in \(\mathcal{O}\)\((n^{3})\) time. Note that when calculating the similarity, we perform content matching (e.g. key API-call operation, instruction operands, etc) for each node. Although it does not perfectly prove that the deobfuscated semantics and the original semantics are completely equivalent, we believe that matching node content can further enhance the original similarity comparison algorithm, thus ensuring that the similarity value can reflect the accuracy of deobfuscation results to a certain extent.

We use the formula below to calculate the CFG similarity:
\begin{equation}
Sim(G_1,G_2) = 1-\frac{\sigma(G_1,G_2)}{|N_1|+|N_2|+|E_1|+|E_2|}\label{CFGSim}
\end{equation}

Here, \textbf{\(\sigma(G_1,G_2)\)} represents the edit distance between \(G_1\) and \(G_2\), \(|N_i|\) is the number of nodes in \(G_i\), and \(|E_i|\) is the number of edges in \(G_i\). The score given by Pearson correlation~\cite{pearson1896mathematical} is in the range of [-1,1]. The closer the score is to 1, the more similar are the structures of these two graphs. A similarity score less than 0 means that two graphs are completely dissimilar~\cite{chan2014method}.

% Previous work suggested that after obfuscation, the CFG similarity score is roughly 0.2, while
% the CFG similarity score achieves higher than 0.8 after deobfuscation is a promising result.

\subsubsection{Semantic Equivalence}
\label{sec:semantic}
\mbox{} \\ 
Though we adopt strict block matching in the calculation of CFG similarity, it is not convincing enough to prove that the deobfuscated function is semantically equivalent to the original one. 
%%In other word, even a CFG similarity value is 1.0, we should also prove the correctness of our deobfuscation result. 
Due to the change of the offset and function size, binary rewriting needs to be performed on each subsequent instruction and function from the $*.so$ file, which may bring disturbances to the deobfuscation result. Also, to the best of our knowledge, there is no effective and accurate way to solve the problem of binary rewriting yet, even from the intermediate representation level. To minimize the introduction of bias caused by other tools, we choose to show our deobfuscation result of the CFF method on the disassembly instruction level, which is useful for reverse analysts to analyze malicious apps.

Regarding correctness, after applying deobfuscation, we use Angr's bit vector to generate 1,500 concrete value inputs (the 500 smallest integers, the 500 largest ones, 500 random others) and test whether the two corresponding outputs (origin and deobfuscated) are identical. If yes, we consider the deobfuscated code as semantically equivalent to the original one. We use \texttt{Input/Output equivalence (I/O)} to represent this evaluation process, which has also been widely used in previous work\cite{salwan2018symbolic}. In order to visualize the experimental results, we use the percentage of the identical I/O in 1500 times of experiments as a quantitative indicator.

% 这里考虑两种方法，一是比较每个基本块的匹配程度，二是比较同样输入的输出的一致性。

%  with 1 indicating total positive correlation, 0 indicating no correlation, and -1 indicating negative correlation. In another word, 
% After completing the comprehensive deobfuscation analysis on the open source C benchmark, we look into the deobfuscating accuracy of our approach on Open Source Android Native Code. We evaluate the accuracy of our deobfuscating approach of BCF and CFF on functions in these binaries.

\subsection{\textbf{Evaluation on C/C++ Obfuscation Benchmarks}}

% \haoyu{First, detailed describe the dataset.}

% \haoyu{for each pass, (1) one paragraph to summarize the result (2) one paragraph to describe one case (better to show the good one) (3) one paragraph to explain the outliers}

Because InsSub only works on computation-intensive programs, we select 5 programs with many computation-related instructions from the benchmark to evaluate the result of InsSub deobfuscation. 

For BCF and CFF, we apply them to all the 94 programs in the benchmark and use \Name to deobfuscate them. We classified the benchmark programs into 5 categories based on their main functionalities: (1) sorting algorithm (Sort), (2) searching algorithm (Search),(3) mathematical calculation (Math), (4) string manipulation (String) and (5) conversion between number and string (Num2String). Furthermore, we also evaluated \Name on cases when all three obfuscation techniques are applied.

% For the source code in the small programs folder, we classify them according to their program structure. In addition, five source codes in the small program folder are too simple (i.e. the unobfuscated function has only one basic block), so we did not make analysis on them. 

\subsubsection{Instruction Substitution}
\mbox{} \\ 
The five programs used in the evaluation include: (1) Binary search, (2) Merge sort, (3) String splicing operation, (4) Quicksort and (5) a program with five arithmetical operations. For these programs, we first apply InsSub to obfuscate the samples and then use \Name to deobfuscate them. Note that for the pass of instruction substitution, we could deobfuscate the results to the binary level.

% The result is shown in Table~\ref{Tab:Benchmark:InsSubtable}. 
\textbf{Euclidean Distance.}
Column 2 and 3 in Table~\ref{Tab:Benchmark:InsSubtable} list the Euclidean distance between the obfuscated function and the unobfuscated one (\(D_{ob}\)), and between the deobfuscated one and the unobfuscated one (\(D_{de}\)), respectively. 
% The smaller the value, the more similar the instructions. 
As we can see, distances between obfuscated functions and unobfuscated ones are in the range of 14 to 44. After deobfuscation, the distances of the recovered functions are all below 10. Note that the gap between deobfuscated and unobfuscated binary (\(D_{de}\)) is mainly caused by the assignment of registers and memory addresses, e.g., \texttt{move}, \texttt{load} and \texttt{store} operations. 
% , when there are relatively more calculation operations in the source code, the obfuscated and the unobfuscated binary have a relatively larger gap at the instruction complexity\haoyu{suggest cut this sentence}. 

\textbf{The number of obfuscated operations.}
We also manually checked the obfuscated and deobfuscated programs to measure the number of obfuscated operations (\(N_{ob}\)) introduced and the number of operations (\(N_{de}\)) successfully recovered, as listed in Column 4 and Column 5 in Table~\ref{Tab:Benchmark:InsSubtable}. All the obfuscated operations are successfully recovered to binary files in our experiment. 

\textbf{Semantic Equivalence.}
We then applied the aforementioned approach (Section~\ref{sec:semantic}) to evaluate the semantic equivalence of deobfuscated binaries and the original ones. As shown in the \textit{``SE''} column of Table~\ref{Tab:Benchmark:InsSubtable}, for each program, we get identical outputs for all the 1500 generated inputs, suggesting that the deobfuscated binaries are semantically equivalent to the original ones.
%%Also the deobfuscated binary files are proved to be semantic equivalent to the original ones by the previous mentioned \texttt{I/O} approach(Column \textit{``SE''}).
% In addition, the function of the recovered binary is also preserved intact\haoyu{To be modified}.
% Also we verified the correctness of the function recovery
%   And the gap between deobfuscated and unobfuscated binary is mainly caused by the assignment of registers and memory addresses, such as move and load operation. 

\begin{table}[t]
    \centering
%   \fontsize{6.5}{8}\selectfont

  \begin{threeparttable}
  \caption{InsSub deobfuscation result.}
  \vspace{-0.1in}
  \label{Tab:Benchmark:InsSubtable}
    \begin{tabular}{c|ccccc}
    \toprule
    Name&\(D_{ob}\) & \(D_{de}\) & \(N_{ob}\) & \(N_{de}\) & SE
    \cr
    \midrule
    binarysearch & 17 & 9 & 5 & 5 & 100\% \cr
    mergesort & 16 & 8 & 2 & 2 & 100\% \cr
    concatstrings & 22 & 0 & 3 & 3 & 100\% \cr
    quicksort & 14 & 3 & 2 & 2 & 100\% \cr
    basic\_arithmetic\_operators & 44 & 10 & 8 & 8 & 100\% \cr
   \bottomrule
    \end{tabular}
      \vspace{-0.2in}
    \end{threeparttable}

\end{table}

\subsubsection{Bogus Control Flow}
\mbox{} \\ 
BCF offers two additional obfuscation options \texttt{bcf\_loop} and \texttt{bcf \\ \_prob}, which controls the obfuscation times (default 1) and obfuscation intense (default 30\%).
To evaluate the effectiveness of \Name on BCF deobfuscation, we applied two different levels of obfuscation intense: the default obfuscation (bcf\_prob=30\%), and enhanced obfuscation (bcf\_prob=50\%). We are also able to recover the deobfuscated code to the binary level.
% They are situations of default and bcf\_prob=50\% respectively. 

\textbf{Result.}
Table~\ref{table:opencresult} shows the overall result. 
Column \textit{``BCF''} and Column \textit{``BCF\_re''} shows the result at the default obfuscation level, and Column \textit{``BCF\_50\%''} and Column \textit{``BCF\_50\%\_re''} shows the result of the enhanced one. 
At the default level, the CFG similarity between the original program and the obfuscated program is 0.453 on average, while the CFG similarity could reach up to an average of 0.870 after the deobfuscation process. 
For the enhanced obfuscation, CFG similarity score of most obfuscated functions has a significant decrease, with an average score of 0.296. Nevertheless, our system could achieve similar good results at different obfuscation level, with an average CFG similarity score of 0.855.

{\begin{table*}[ht!]

\caption{Deobfuscation results on the C/C++ Benchmark.}
%%\vspace{-0.1in}
    \centering
  \fontsize{6.5}{8}\selectfont
\resizebox{1\linewidth}{!}{
  \begin{threeparttable}
\renewcommand{\arraystretch}{} % Default value: 1  
    \begin{tabular}{c|ccc|ccc|ccc|ccc|ccc}
    \toprule
    Name & BCF & BCF\_re  & SE & BCF\_50\% & BCF\_50\%\_re & SE & CFF & CFF\_re & SE & CFF-3 & CFF-3\_re & SE & ALL & ALL\_re & SE 
    \cr
    \midrule

selectionsort & 0.904 & 0.938 & 100\% & 0.398 & 0.938 & 100\% & 0.151 & 0.910 & 100\% & -0.160 & 0.875 & 100\% & -0.128 & 0.747 & 100\% \cr
 
bubblesort & 0.709 & 0.852& 100\%  & 0.469 & 0.970& 100\%  & 0.200 & 0.818& 100\%  & -0.160 & 0.726& 100\%  & -0.230 & 0.821& 100\%  \cr
 \midrule
 
binarysearch & 0.759 & 0.897& 100\%  & 0.372 & 0.923 & 100\% & 0.137 & 0.868 & 100\% & -0.192 & 0.868 & 100\% & 0.023 & 0.816 & 100\% \cr
 
binarysearchrec & 0.597 & 0.750 & 100\% & 0.326 & 0.750 & 100\% & 0.318 & 0.843 & 100\% & -0.130 & 0.816 & 100\% & 0.023 & 0.816 & 100\% \cr
 \midrule
 
gcd & 0.660 & 0.872 & 100\% & 0.388 & 0.872 & 100\% & 0.075 & 0.776 & 100\% & -0.211 & 0.776 & 100\% & -0.294 & 0.776 & 100\% \cr
 
 lcm & 0.152 & 0.912 & 100\% & 0.365 & 1.000 & 100\% & 0.172 & 0.857 & 100\% & -0.218 & 0.912 & 100\% & -0.218 & 0.912 & 100\% \cr
 \midrule
 
 concatstrings & 0.213 & 0.769 & 100\% & 0.133 & 0.795 & 100\% & 0.213 & 0.778 & 100\% & -0.261 & 0.872 & 100\% & -0.124 & 0.872 & 100\% \cr
 
 reverse & 0.280 & 1.000 & 100\% & 0.370 & 0.900 & 100\% & 0.292 & 0.926 & 100\% & -0.252 & 0.828 & 100\% & -0.114 & 0.828 & 100\% \cr
 \midrule
 
 romannumerals & 0.234 & 0.797 & 100\% & 0.321 & 0.816 & 100\% & 0.164 & 0.797 & 100\% & -0.250 & 0.869 & 100\% & -0.340 & 0.807 & 81\% \cr
 
 decimaltobinary & 0.184 & 0.906 & 100\% & 0.253 & 0.783 & 100\% & 0.310 & 0.854 & 100\%  & -0.174 & 0.854 & 100\% & 0.016 & 0.854 & 100\% \cr
 \midrule
 
 bkdrhash & 0.175 & 0.828 & 100\% & 0.155 & 0.822& 100\% & 0.294 & 0.926 & 100\% & -0.165 & 0.828 & 100\% & -0.070 & 0.926 & 67\% \cr 
 
 djbhash & 0.378 & 0.828 & 100\% & 0.409 & 0.813 & 100\% & 0.264 & 0.926 & 100\% & -0.173 & 0.926 & 100\% & -0.198 & 0.742 & 100\% \cr
 \midrule
 
 \begin{tabular}[c]{@{}c@{}}1b-1-2-2-1-\\ gt127-0-0-0\end{tabular} & 0.618 & 1.000 & 100\% & -0.029 & 1.000 & 100\% & 0.304 & 1.000 & 100\% & -0.135 & 1.000 & 100\% & -0.126 & 1.000& 100\%  \cr
 \midrule
 
 \begin{tabular}[c]{@{}c@{}}1b-1-2-1-1-gtsum\\ 127\_dep-1-0-1\end{tabular} & 0.618 & 0.893 & 100\% & 0.083 & 0.737 & 100\% & 0.143 & 0.821 & 100\% & -0.224 & 0.750 & 100\% & -0.223 & 0.821 & 100\% \cr
 \midrule
 
 \begin{tabular}[c]{@{}c@{}}16b-1-1-0-0\\ -dc-2-2-0\end{tabular} & 0.382 & 0.900 & 100\% & 0.534 & 0.821 & 100\% & 0.475 & 0.722 & 100\% & -0.161 & 0.722 & 100\% & -0.052 & 0.788 & 100\% \cr
 \midrule
 
 \begin{tabular}[c]{@{}c@{}}1b-4-2-0-0\\ -dc-2-2-0\end{tabular} & 0.618 & 1.000 & 100\% & 0.196 & 0.742 & 100\% & 0.216 & 0.821 & 100\% & -0.149 & 0.821 & 100\% & -0.242 & 0.821 & 100\% \cr
 \midrule
 \midrule
 \begin{tabular}[c]{@{}c@{}}Overall similarity \\ of Approach $\triangle$ \end{tabular} & - & - & - & - & - & - & 0.206 & 0.674 & -  & -0.185 & 0.393 & - & -0.132 & 0.450 & -  \cr
 \midrule
  
 \begin{tabular}[c]{@{}c@{}}Overall similarity \\ of \Name \end{tabular} & 0.453 & 0.870 & 100\% & 0.296 & 0.855 & 100\% & 0.206 & 0.807 & 100\%  & -0.185 & 0.722 & 90\% & -0.132 & 0.734 & 94\%  \cr

   \bottomrule
    \end{tabular}
    \end{threeparttable}
}
 \vspace{-0.1in}
  \label{table:opencresult}
\end{table*}
}

\textbf{Semantic Equivalence.}
%%Although \Name could deobfuscates the BCF approach to binary level, we opt to use I/O approach to automatically evaluate our deobfuscation results are semantic equivalent to the original ones.
As shown in Table~\ref{table:opencresult} (cf. Column ``SE''),
all deobfuscation results are equivalent to the original ones, for the two levels of BCF obfuscation. 
We further manually checked the deobfuscation results of the 16 functions shown in Table~\ref{table:opencresult} by analyzing the instructions in each block of the CFGs, and found that their semantics are indeed equivalent to the original ones.

\subsubsection{Control Flow Flattening}

% Then we moved our line of sight to the deobfuscation of Control Flow Flatten. 
% \haoyu{rewrite this paragraph with the following texts, just describe what does it mean.}

\mbox{} \\ 
CFF offers an additional obfuscating option \texttt{split\_num}, which means to split the original basic block a specified number of times, to increase the complexity of obfuscated control flow.
In the evaluation, we have applied two different obfuscation levels: the default CFF obfuscation\footnote{O-LLVM itself does not split the basic block. However, due to optimizations during ARM compilation, basic block splitting still exists to some extent.}, and an enhanced obfuscation level with the basic block splitting option on \texttt{split\_num=3}.
% \haoyu{rewrite}. 
% The results of these two situation are shown in column 'CFF' and 'CFF-3'. And the column name add '\_re' is the deobf. similarity score of that situation.

% , as a result, the deobfuscationg work for this pass is also the focus of our evaluation. 

% We started from deobfuscation work on open source C benchmark with control flow flatten at default setting. 

% \haoyu{Overall results here.} 

\textbf{Result.} Table \ref{table:opencresult} shows the results. Column \textit{``CFF''} and Column \textit{``CFF\_re''} shows the results on default obfuscation of CFF, and Column \textit{``CFF-3''} and Column \textit{``CFF-3\_re''} shows the results of enhanced obfuscation, respectively. 
Obviously, with CFF obfuscation, the obfuscated code achieved significant differences compared to original ones considering the CFG similarity. 
At the default obfuscation level, the average CFG similarity score is only 0.206 after obfuscation, while our deobfuscation results could achieve a similarity score of 0.807 on average. After activating basic block splitting, similarity scores between the obfuscated and original function are almost all negative correlation (-0.185 on average), which indicates that the obfuscated CFG is completely different from the original one. The listed deobfuscation result of the enhanced CFF ranges from 0.72 to 1. The average value on all 90 functions(4 failures with O-LLVM compiling) is 0.722.

% \haoyu{One paragraph here to clearly describe the example! Merge all the following TEXTS!}

% While CFF obfuscating approach being used, the function produces significant differences compare to the unobfuscated one. By default obfuscate setting, the average score of CFG similarity is 0.205562, and some of them are even lower than 0.1. And our deobfuscation result got an average score of 0.806897 on a total number of 94 functions. After activating basic block split mode, similarity scores between the obfus. and unobfus. function are almost all negative correlation, which indicates that the obfuscated CFG is completely different from the unobfuscated one. 

% \begin{figure*}[htbp]
% \centering
% \begin{minipage}[t]{0.15\linewidth}
% \includegraphics[width=\linewidth]{evaluation_pic/16-origin.png}
% \label{fig:16-origin}
% \caption{Unobf.}
% \end{minipage}%
% % \begin{minipage}[t]{0.2\linewidth}
% % \includegraphics[width=\linewidth]{pictures/16-1.png}
% % \label{fig:side:b}
% % \end{minipage}
% \begin{minipage}[t]{0.5\linewidth}
% \includegraphics[width=\linewidth]{pictures/16-3.png}
% \label{fig:16-3}
% \caption{Obf.}
% \end{minipage}
% \begin{minipage}[t]{0.25\linewidth}
% \includegraphics[width=\linewidth]{evaluation_pic/16-re.png}
% \caption{Deobf.}
% \label{fig:16-re}
% \end{minipage}
% \end{figure*}

% As the data shows, except a few functions, our system achieves close results on both cases of a same function. 
% We first use program 16b-1-1-0-0-dc-2-2-0 as example to illustrate the reason why gaps occur between the Deobf. and Obf. function.

\begin{figure}[t]
\centering
\includegraphics[width=0.5\textwidth]{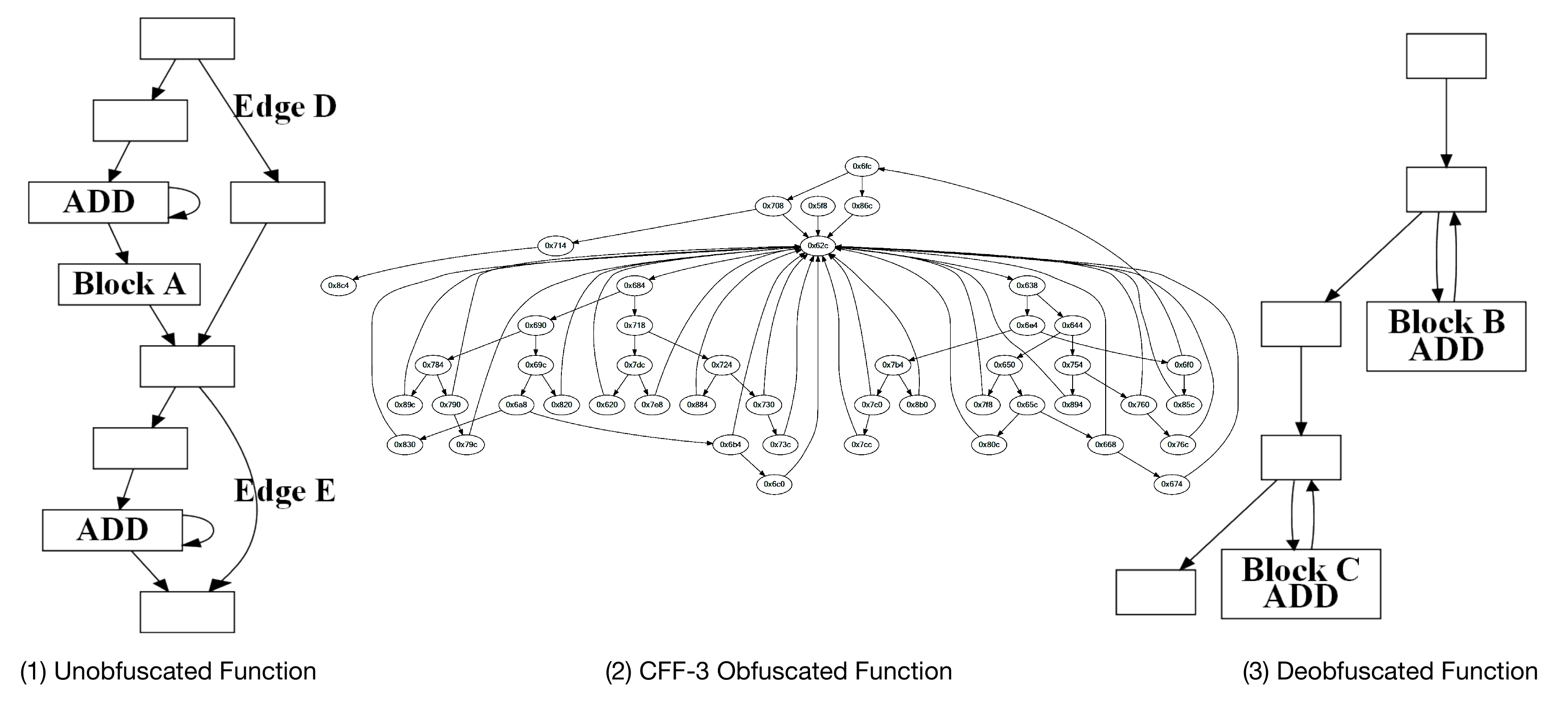}
\vspace{-0.15in}
\caption{A Deobfuscation Example of CFF.}
\label{fig:16}
\vspace{-0.1in}
\end{figure}

\textbf{Case Study.} Experimental results suggest that our approach could achieve promising results on most cases. However, for several cases, the CFG similarity of deobfuscated code with the original binary is roughly 0.7. By further analyzing these cases, we found that the main reason is the redistribution of semantics in the original basic blocks. We use the program 16b-1-1-0-0-dc-2-2-0 as an example to illustrate this situation.
As shown in Fig. \ref{fig:16}. This is a function that calculates twice the sum of the ASCII codes of the input string. 
% To clearly show our deobfuscated result, we display the key instructions of each basic block in the CFG. 
The program got a recover result of 0.722 in both default and enhanced obfuscation. 
% The recovered CFG is shown in Fig. \ref{fig:16}. 
Block A in the original CFG (Fig. \ref{fig:16} (a)) is a transition block, and we assume that its parent and child node are directly connected. Also, the two self-loop basic blocks in the original CFG become the loop between two blocks in the recovered result. The key operation `ADD' of the two self-loop structures in the original function are now implemented in the split-out blocks B and C, as shown in Fig. \ref{fig:16} (c). Thus operations on edges D and E can be merged to another branch of its starting node.
This example suggests that CFF may separate constraints and operations during obfuscation, which will lead to re-matching of nodes and edges in the recovered result. 
However, this scenario will not affect the analyzing results, but instead simplifying the analysis process. 

\textbf{Semantic Equivalence.} Here we also use the \texttt{I/O} approach to check whether our deobfuscation results are semantic equivalent to the original ones.
As shown in Table~\ref{table:opencresult} (cf. Column \textit{``SE''}), for the default CFF obfuscation, our deobfuscation results are totally identical with the original semantics.
However, for the enhanced CFF obfuscation, we found some cases shared different outputs with original ones, thus the overall semantic equivalent evaluation result is 90\%. 
By analyzing these exceptional cases, we found that the main reason is introduced by function splitting (cf. Figure~\ref{fig:uninit}), as we skipped the \texttt{BL and BLX} instructions when we perform the semantic equivalence evaluation.
For such exceptional cases, we manually compared the deobfuscated programs and the original ones, and found that their semantics should be identical if we consider the \texttt{BL and BLX} instructions (superimposing the recovered functions), which we will elaborate in both following subsection and Section~\ref{sec:discussion}

\subsubsection{Full Obfuscation (All 3 Techniques)}
\mbox{} \\ 
The aforementioned results suggest that \Name can recover the obfuscated samples accurately for each pass of O-LLVM. We further evaluate it if three obfuscation techniques are used at the same time.

% After completing the separate analysis of the two obfuscation methods, we implement our deobfuscation system on the mixture situation of O-LLVM's three passes. 

\textbf{Result.} The result is shown in the last three Columns in Table \ref{table:opencresult}. Obviously, the combined techniques introduce strong obfuscation impacts, with negative CFG similarity scores for most cases. The average CFG similarity score after obfuscation is -0.132 on average. However, \Name could achieve good results compared with single pass evaluation, with an average CFG similarity score of 0.734 after deobfuscation. For example, the \textit{romannumberals} function, which converts decimal numbers to Roman characters, has more than 500 basic blocks after hybrid obfuscation of O-LLVM. The CFG similarity score of the obfuscated function is -0.340, which is the lowest. The deobfuscating result on \textit{romannumberals} is 0.807, showing that \Name is resilient to complicated code obfuscation.
% , even though we have no prior knowledge about which techniques are used.

\textbf{Semantic Equivalence.} 
We further perform semantic equivalence evaluation.
As shown in the last column of Table~\ref{table:opencresult}, besides several exceptional cases (e.g., bkdrhash), we could get identical outputs for most cases, and thus the final average result is 94\%. Note that we further manually analyzed the extreme cases, and found the leading reason is the same with the cases we identified in CFF deobfuscation, i.e., function splitting (cf. Section~\ref{sec:discussion}).

%%The \texttt{I/O} approach is also used here to automatically prove deobfuscation results are semantic equivalent to the original ones. As Column \textit{``SE''} in Table~\ref{table:opencresult} shows, all listed 16 deobfuscation results for full obfuscation on C/C++ obfuscation benchmark are completely equivalent to the original semantics. The \texttt{I/O} percentage of semantic equivalent samples in total is 94\%. We also manually analyzed the deobfuscation results of the 16 functions shown in Table~\ref{table:opencresult}, and proved that their semantics are completely equivalent to the original semantics. 

\subsubsection{Comparing with Existing Studies.}
\mbox{} \\ 
%In previous studies, Francis Gabriel~\cite{gabriel2014deobfuscation} attempted to recover O-LLVM obfuscated code on the X86 platform. His work is based on the Miasm~\cite{miasm} framework to reverse the obfuscated function only to a CFG in all three kind of obfuscation approaches in O-LLVM. However, as we illustrated in Section \ref{sec:background}, it did not tackle multiple challenges introduced by ARM and Android. 
In previous studies, Francis Gabriel~\cite{gabriel2014deobfuscation} attempted to recover O-LLVM obfuscated code on the X86 platform based on the Miasm~\cite{miasm} framework. Specifically, his work only aimed to recover corresponding CFGs of functions protected by one of the three obfuscating techniques in O-LLVM. However, as we illustrated in Section \ref{sec:background}, it did not tackle multiple challenges introduced by ARM and Android. 

In this subsection, we would like to compare his work (we use the symbol $\triangle$ to represent it) with ours. Because the Approach $\triangle$ could not be used to deobfuscate Android native functions directly, we make the compromise to remove the context inheritance algorithm and control flow reconstructing rules in \Name and assume that Approach $\triangle$ could achieve the alike deobfuscation result with the degraded \Name. Obviously it is unfair to our approach considering the adoption of numerous improvements and innovations beside these two dominating ideas. However, if \Name could produce better results than Approach $\triangle$ in this scenario, the deobfuscating ability of \Name will be convincing. 
% not basic, it's ACE idea, lol

As shown in Table \ref{table:opencresult}, not surprisingly, the average similarities of the obfuscated functions in these three cases (CFF, CFF-3 and ALL) are exactly the same. However, the average deobfuscated CFG similarity of Approach $\triangle$ in the CFF situation is 0.674. Even worse, the similarity values decline dramatically to 0.393 and 0.450, CFF-3 and ALL respectively. With regard to those relatively large or basic-block-split functions, our further manual investigation demonstrates that the recovered result of Approach $\triangle$ is fragmented, and the corresponding caller-callee relationships between many basic blocks are lost, i.e., there does not exist any recovered path to reach the Return block. Due to its essence of unreliable results, we do not use the \texttt{I/O} approach to prove the correctness of Approach $\triangle$.
As such, we have demonstrated that \Name could achieve more accurate and convincing deobfuscation results.

\begin{table}[t]
\centering
\caption{Deobfuscation result on open-source Android apps.}
%%\vspace{-1ex}
\resizebox{1\linewidth}{!}{

  \begin{threeparttable}

    \begin{tabular}{c|c|ccc|ccc}
    \toprule
    APK Name & Function Name & CFF & CFF\_re & SE & BCF & BCF\_re & SE 
    \cr
    \midrule
 & rdft & 0.199 & 0.919 & 100\% & 0.166 & 0.702  & 100\%\cr 
 & newFFT & 0.174 & 0.792 & 100\% & 0.704 & 0.792  & 100\%\cr 
 & makect & 0.325 & 0.913 & 100\% & 0.244 & 0.786  & 100\%\cr 
 \multirow{-4}{*}{Practice Hub} & Total  10 & 0.365 & 0.843 & 100\% & 0.446 & 0.795  & 100\%\cr
 \midrule
  
  & \begin{tabular}[c]{@{}c@{}}free\end{tabular} & 0.228 & 0.889  & 100\%& 0.338 & 0.752  & 100\%\cr
 
 & \begin{tabular}[c]{@{}c@{}}seekToFrame\end{tabular} & 0.017 & 0.725 & 100\% & 0.537 & 0.795  & 100\%\cr
 
 & \begin{tabular}[c]{@{}c@{}}seekToTime\end{tabular} & -0.004 & 0.698  & 100\%& 0.607 & 0.871  & 100\%\cr

\multirow{-4}{*}{\begin{tabular}[c]{@{}c@{}}Overchan\end{tabular}}  & Total 17 & 0.138 & 0.745 & 94\% & 0.339 & 0.761 & 100\% \cr
 \midrule
%  fillPixelsInRowNative
% getAsciiValuesBWNative
%getAsciiValuesWithColorNative
  & \begin{tabular}[c]{@{}c@{}}InRowNative\end{tabular} & 0.389 & 0.717 & 100\% & 0.457 & 0.654 & 100\% \cr
 & \begin{tabular}[c]{@{}c@{}}BWNative\end{tabular} & 0.417 & 0.714 & 100\% & 0.683 & 0.892 & 100\% \cr
 & \begin{tabular}[c]{@{}c@{}}ColorNative\end{tabular} & 0.009 & 0.712 & 100\% & 0.486 & 0.712 & 100\% \cr 
\multirow{-4}{*}{\begin{tabular}[c]{@{}c@{}}AsciiCam\end{tabular}} & Total  3 & 0.266 & 0.714 & 100\% & 0.542 & 0.753 & 100\% \cr
 \midrule
  
 & anagrams\_init & 0.133 & 0.802 & 100\% & 0.143 & 0.911 & 100\%\cr  
  
 & \begin{tabular}[c]{@{}c@{}}Anagrams\_uninit\end{tabular} & \begin{tabular}[c]{@{}c@{}}0.791
(0.077)\end{tabular} & \begin{tabular}[c]{@{}c@{}}0.333
(0.875)\end{tabular}  & 27(100)\% & \begin{tabular}[c]{@{}c@{}}0.800
(0.122)\end{tabular} & \begin{tabular}[c]{@{}c@{}}0.160
(0.810)\end{tabular}  & 21(100)\%\cr 
  
  & anagram & 0.174 & 0.830 & 100\% & 0.058 & 0.784 & 100\%\cr

\multirow{-4}{*}{\begin{tabular}[c]{@{}c@{}}Agram\end{tabular}}  & Total 24 & 0.198 & 0.761 & 100\% & 0.511 & 0.777 & 100\% \cr 
 \midrule
 & createWnnWork & 0.067 & 0.747 & 100\% & 0.306 & 0.848  & 100\%\cr 
 & freeWnnWork & 0.149 & 0.875 & 100\% & 0.316 & 0.737  & 100\%\cr 
\multirow{-3}{*}{NicoWnnG} & Total 2 & 0.108 & 0.811  & 100\%& 0.311 & 0.792  & 100\%\cr 
   \bottomrule
    \end{tabular}
    \end{threeparttable}
    
}
\vspace{-4ex}
\label{table:mobileresult}
\end{table}

\subsection{\textbf{Evaluation on Open-Source Android Apps}}

The five open source Android apps used in our evaluation are: (1) Practice Hub (com.proch.practicehub.src), a tool for musicians; (2) Overchan (bus.chio.wishmaster), an app for browsing different kinds of imageboards; (3) AsciiCam (com.dozing catsoftware.asciicam), a photography app generates ASCII images in real time, with more than 100K downloads in Google Play; (4) Agram (us.achromatic metaphor.agram), a tool to list single-word and multi-word anagrams in English; (5) NicoWnnG (net.gorry.android.input.nicownng), a keyboard IME with more than 50K downloads in Google Play.

\textbf{Result.} For each open-source app, we first extract its native source code. Then we use O-LLVM to perform obfuscation when building $*.so$ according to Application.mk and Android.mk in the original Android projects.
% for obfuscation compilation of these native code according to the original Application.mk and Android.mk in their project folders\haoyu{rewrite}. 
In the deofuscation process, we filtered the functions that contain only one basic block. At last, we analyzed 56 functions in total. 
% Because the previous result did not show a big difference under different strength of obfuscation, we analyzed all functions that obfuscated by default CFF and BCF separately.

% using our deobfuscation system and then used O-LLVM for obfuscation compilation of these native code according to the original Application.mk and Android.mk in their project folders.

Table \ref{table:mobileresult} shows the overall deobfuscation results. 
For binaries obfuscated with BCF (with average CFG similarity score ranging from 0.311 to 0.542), our approach could achieve a CFG similarity score in the range from 0.753 to 0.795 on average after deobfuscation. For binaries obfuscated with CFF (with average CFG similarity score ranging from 0.108 to 0.365), our approach could achieve a CFG similarity score around 0.75 on average after deobfuscation. Although the overall result is acceptable, for a few cases, our approach did not achieve good results as other cases. We further manually analyzed these cases.

{\begin{figure}[t]
\includegraphics[width=\columnwidth]{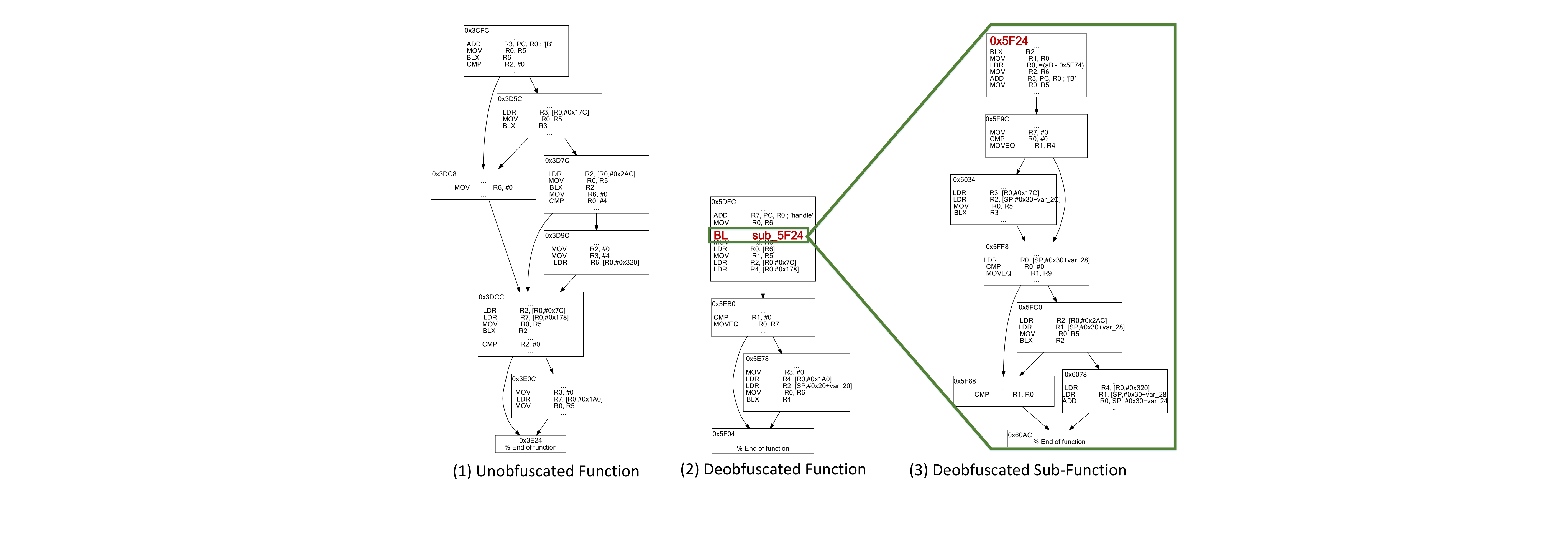}
%%\vspace{-0.2in}
\caption{Another deobfuscation case (Anagrams\_uninit).}
\label{fig:uninit}
%%\vspace{-0.1in}
\end{figure}}

\begin{figure*}[t]
\includegraphics[width=\linewidth]{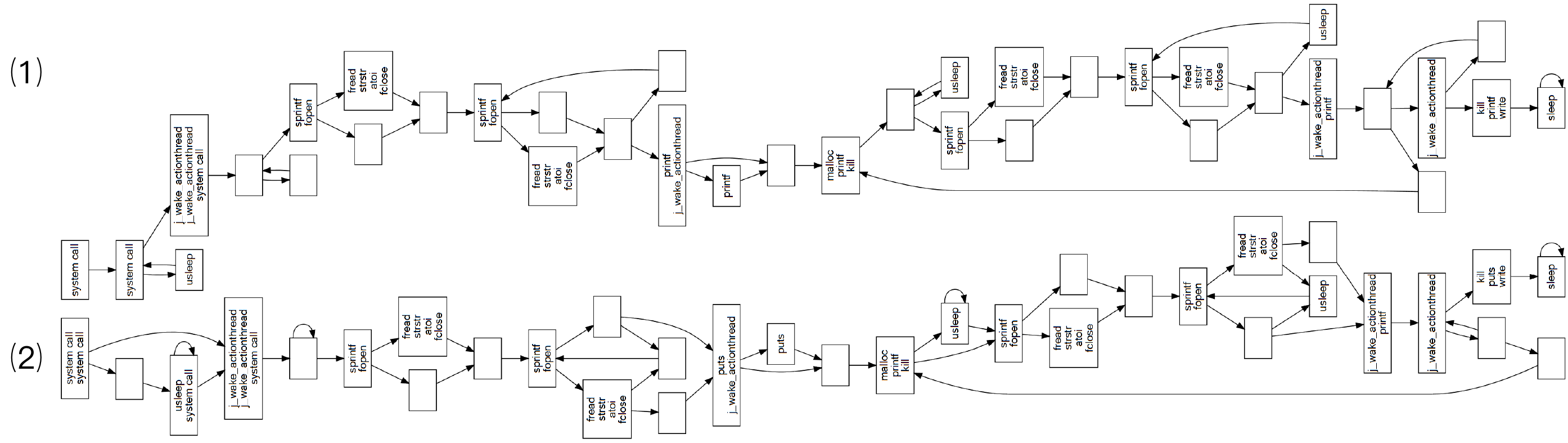}
\caption{CFG comparison of search\_goodnum: ((1) Deobfuscated Graph; (2) Unobfuscated Graph. \textit{Obfuscated Graph in Fig. ~\ref{fig:motivating}.})}
\label{fig:goodnum}
\vspace{-0.1in}
\end{figure*}

We found that the main reason leading to the CFG inconsistency is \emph{function splitting}. We use function \textit{``Anagrams\_uninit''} in \textit{``Agram''} to illustrate it. Sometimes an operation in the original $*.so$ can be implemented by one function, while in the obfuscated binary, part of this operation may be split out and used as a sub-function due to optimizations. As shown in Fig. \ref{fig:uninit} (2), the CFG of deobfuscated function only contains 4 basic blocks. In the first block, this recovered function calls \textit{``sub\_5F24''}, which is also obfuscated by O-LLVM. The deobfuscation of \textit{``sub\_5F24''} is shown in Fig. \ref{fig:uninit} (3). After superimposing these two recovered functions' CFG, it can be seen that the similarity between original and recovered CFG is very high. The deobfuscation results are 0.875 for CFF and 0.810 for BCF, compare to 0.333 and 0.16 respectively. This result suggests that function splitting could greatly affect the result. Thus, in our implementation, we take the function splitting into consideration, which is not carefully considered in previous studies~\cite{D-OLLVM2017,gabriel2014deobfuscation,yadegari2015generic}.

\textbf{Semantic Equivalence.} As shown in the Column \textit{``SE''} of Table~\ref{table:mobileresult}, except for app ``Overchan", all the recovered functions of the rest apps get 100\% identical outputs in our \texttt{I/O} approach. We further manually analyzed such cases, and found that their semantics are completely equivalent to the original semantics, while this result is introduce by the limitation of our I/O approach (cf. Section~\ref{sec:discussion}).

\begin{table}[t]
    \centering
  \fontsize{6.5}{8}\selectfont

  \begin{threeparttable}
  \caption{The deobfuscation result on the PoC implementation of towelroot.}
  \label{CVEtable}
%%  \vspace{-0.15in}
    \begin{tabular}{c|ccc|ccc}
    \toprule
    function name & CFF & CFF\_re & SE & ALL & ALL\_re & SE
    \cr
    \midrule
    accept\_socket & 0.302 & 0.797 & 100\% & -0.270 & 0.913 & 100\%  \cr 
 make\_socket &  0.346 &  0.667& 100\% &  -0.267 &  0.935& 100\%  \cr  
init\_exploit & 0.203 & 0.903& 100\% & -0.044 & 1.000& 100\% \cr 
 send\_magicmsg &  0.149 &  0.963& 100\% &  0.149 &  0.962& 100\%  \cr 
read\_pipe & 0.111 & 1.000 & 100\%& -0.265 & 1.000& 100\%  \cr  
 search\_goodnum &  0.193 &  0.939 & 100\%&  -0.245 &  0.900& 100\%  \cr 
make\_action & 0.367 & 1.000& 100\% & -0.157 & 0.850& 100\%  \cr  
 write\_kernel &  0.211 &  0.966& 100\% &  -0.210 &  0.703 & 100\% \cr 
wake\_actionthread & 0.216 & 0.927& 100\% & -0.220 & 0.912& 100\%  \cr 
 main &  0.194 &  1.000 & 100\% &  -0.253 &  0.652 & 100\% \cr 
write\_pipe & 0.111 & 1.000 & 100\% & -0.161 & 1.000 & 100\% \cr
   \bottomrule
    \end{tabular}
    \end{threeparttable}
    \vspace{-0.2in}
\end{table}

\subsection{Evaluation on Real-world Exploits}
% In this subsection, we would like to focus on Android Root Exploits (aka REs), a sort of binary tool which can be used to obtain the root privilege by exploiting privilege escalation vulnerabilities. 
% The use of REs was full of controversial~\cite{gasparis2017detecting}. On one side, as the key part of some legitimate products provided by root providers, REs have been used to serve normal customers for legal use (rooting, as well as jailbreak); on the other side, REs have already been embedded by malware in the wild to compromise the target Android devices. According to our observation, many embedded REs were obfuscated for various purposes, which became a big obstacle for security researchers.  

CVE-2014-3153~\cite{CVE2014} is a well-known generic vulnerabilities in the Linux kernel, i.e., the Fast User space muTEX (futex) subsystem, which is the basis of several mutual exclusion mechanisms. 
Its exploitation (e.g., towelroot~\cite{towelroot} and different variants) had been widely spread around world. The original towelroot was protected by O-LLVM, which took researchers a long time to understand the technical details by performing laborious reverse engineering. As shown in the motivating example (Fig.~\ref{fig:motivating}), it is quite difficult, if not impossible, for researchers (even experienced experts) to figure out the details from the obfuscated CFG directly.

% The complexity comes from the essence of a race condition exploit with multiple threads, where the sequence of invoking system calls becomes the key to understand the subtle logic. 

\textbf{Result.} As of this writing, the source code of the original towelroot is still unavailable. Fortunately, some researchers have published their results, which can be regarded as the identical PoCs of the original one. We will then demonstrate the capability of our system based on one of them~\cite{git-cve}, and the results are summarized in Table \ref{CVEtable}. For simplicity, we focus on two key functions, i.e., ``search\_goodnum'' and ``send\_magicmsg'', to illustrate our work. 

% Looks like we do not need CFF results here.

The deobfuscation results of ``send\_magicmsg'' and ``search\_good-num'' in ``CFF'' situation are $0.963$ and $0.939$ respectively, which means that the deobfuscated CFGs are almost the same as the original ones. In ``ALL'' obfuscation situation, these two value are $0.962$ and $0.900$ respectively. Fig. ~\ref{fig:goodnum} shows the comparison between the two CFGs of ``search\_goodnum''. Compared to the obfuscated CFG shown in Fig.~\ref{fig:motivating}, it is obviously much easier to understand the logic of the exploitation, especially the system call sequences and key operands, from the deobfuscated CFG.

\textbf{Semantic Equivalence.} We also use the \texttt{I/O} approach to prove the deobfuscation results are semantic equivalent to the original function. As shown in Table \ref{CVEtable}, for all cases, all the outputs of deobfuscated functions are identical to the outputs of ones, considering the same 1,500 inputs.

\section{Threats to Validity and Discussion}
\label{sec:discussion}
%  But there are also an important requirement to the input of our system. It needs to ensure that the input binary does not obfuscated by other approach that blurs its semantics, e.g., packing. The defenses of packing in a packed shared library have to be overcome before our system being applied.\haoyu{suggest cut this paragraph}

% \subsubsection{Errors in the underlying framework}
% 由于我们的反混淆工作是以函数为单位的，所以函数基本块的识别非常重要。我们的函数识别是基于ida的识别结果的。由于各个disassembler识别的cfg不一样，可能会对恢复结果造成影响。另外由于我们的符号执行引擎是Angr，对于angr无法识别的或者识别存在误差的指令，也可能会造成结果的不准确。

Although the experimental results suggest that \Name achieves good performance in recovering O-LLVM obfuscated functions, our study, however, carries a few threats to validity.

\textbf{Inherent Limitations of IDA Pro and Angr.}
Since parts of our system build upon several state-of-the-art tools, which might introduce inherent limitations. 
As we perform function-level deobfuscation in this work, it is important to recognize the initial address and the basic blocks in each function. However, it is possible that IDA Pro cannot identify the address of a function accurately, which may lead to inaccuracy of our evaluation. In addition, we rely on Angr to perform symbolic execution, while the instructions that cannot be accurately recognized by Angr will likely cause inaccuracies.
These are the limitations inherited from IDA Pro and Angr, although they rarely occur during our experiments.
%%For example, most of the inconsistencies in our semantic equivalence evaluation were introduced by the incorrect identification of instructions.

\textbf{Semantic Equivalence.}
%%In the previous shown results, not all functions in the \texttt{SE} column have a 100\% \texttt{I/O} matching percentage. We will explain this phenomenon in this section. 
Our experiment suggested that, for a small number of cases, there exist inconsistencies in evaluating the semantic equivalence of deobfuscated binaries and the original ones. We further manually explored these cases and pinpoint the following two reasons.

First, as we use symbolic execution to run the function at the IR level, the \texttt{BL} and \texttt{BLX} instructions sometimes may affect the proper execution of the symbolic execution engine. As shown in Fig.~\ref{fig:uninit}, the semantics of function \texttt{sub\_5F24} is split from the original function. At the very beginning, we try to enter the callee but unfortunately, the execution engine crashed there for quite a few times. Thus, when evaluating semantic equivalence, instead of entering the callee function, we skip them due to the uncertainty of the callee's space. 
Based on the manual analysis of such samples, we observe that the semantics of them are completely recovered by \Name. 
It is worth mentioning that we also skip callee functions in the deobfuscation process, but it will not affect the recovered control flow, as the flow is routed by \(V_{routing}\), which will not be modified cross function or through a callee function.

%%. Because control flow relies on \(V_{routing}\). It will not be modified the cross function or through callee function.

%%We also summarized the samples with such problems and did manual analysis on them. We believe that although there are some differences in \texttt{I/O}, the basic semantics of them are still completely recovered by \Name. Please notice that we also skip callee functions in the deobfuscation process, but it will not affect the recovered control flow. Because control flow relies on \(V_{routing}\). It will not be modified the cross function or through callee function.

% dispatcher中对于寄存器的修改
Second, we saved the symbolic state after each time of symbolic execution in Algorithm 1. Actually, the chopped process of symbolic execution of a block may produce multiple different states before a successor. In the evaluation, \Name only saves one single state, and the state will be updated in the following analysis if it is found to be a non-blank state. It makes our evaluation to be quite effective. However, in some exceptional cases, very few paths may not be found due to the loss of necessary states. 
In our work, we make the number of saved states of one single block configurable, i.e., users could increase the number of states to save.
Increasing the number of saved states will improve our analysis results, as the saved variables are all inherited from the pre-executions.
%%We set the number of saved states of one single block optional for users if they think the increased analyzing time is acceptable. 
%%We believe that increasing the number of saved states will only improve our analysis results because the saved variables are all inherited from the pre-executions.

Although our semantic evaluation approach does not report 100\% identical for several cases, our manually efforts confirmed that all such cases were introduced by the aforementioned reasons, and the recovered programs keep the same semantics actually. 

\section{Conclusion}

We have presented a novel approach for deobfuscating Android native code. It uses taint analysis to make semantic-level deobfuscation, and leverages an enhanced flow-sensitive symbolic execution to rebuild the seriously obfuscated control flow. We have implemented our approach in a system called \Name, and demonstrated that \Name could successfully reverse obfuscations performed by O-LLVM with high accuracy. To the best of our knowledge, this is the first work that tackles the problem of Android native code deobfuscation. We believe that our system could become an useful tool for security analysts and researchers to conduct studies including malware detection and program analysis.

\newpage
\balance
% The next two lines define the bibliography style to be used, and the bibliography file.
\bibliographystyle{ACM-Reference-Format}
\bibliography{reference}


\begin{thebibliography}{61}

%%% the Web address.

\ifx \showCODEN    \undefined \def \showCODEN     #1{\unskip}     \fi
\ifx \showDOI      \undefined \def \showDOI       #1{#1}\fi
\ifx \showISBNx    \undefined \def \showISBNx     #1{\unskip}     \fi
\ifx \showISBNxiii \undefined \def \showISBNxiii  #1{\unskip}     \fi
\ifx \showISSN     \undefined \def \showISSN      #1{\unskip}     \fi
\ifx \showLCCN     \undefined \def \showLCCN      #1{\unskip}     \fi
\ifx \shownote     \undefined \def \shownote      #1{#1}          \fi
\ifx \showarticletitle \undefined \def \showarticletitle #1{#1}   \fi
\ifx \showURL      \undefined \def \showURL       {\relax}        \fi
% invisible to TeX
\providecommand\bibfield[2]{#2}
\providecommand\bibinfo[2]{#2}
\providecommand\natexlab[1]{#1}
\providecommand\showeprint[2][]{arXiv:#2}

\bibitem[\protect\citeauthoryear{??}{CVE}{2014}]%
        {CVE2014}
 \bibinfo{year}{2014}\natexlab{}.
\newblock \bibinfo{title}{CVE-2014-3153}.
\newblock
  \bibinfo{howpublished}{https://cve.mitre.org/cgi-bin/cvename.cgi?name=cve-2014-3153}.
\newblock


\bibitem[\protect\citeauthoryear{??}{git}{2014}]%
        {git-cve}
 \bibinfo{year}{2014}\natexlab{}.
\newblock \bibinfo{title}{CVE-2014-3153 aka towelroot}.
\newblock \bibinfo{howpublished}{https://github.com/orenl/CVE-2014-3153}.
\newblock


\bibitem[\protect\citeauthoryear{??}{tow}{2014}]%
        {towelroot}
 \bibinfo{year}{2014}\natexlab{}.
\newblock \bibinfo{title}{towelroot}.
\newblock \bibinfo{howpublished}{https://towelroot.com/}.
\newblock


\bibitem[\protect\citeauthoryear{??}{Tow}{2014}]%
        {Towelroot1}
 \bibinfo{year}{2014}\natexlab{}.
\newblock \bibinfo{title}{Towelroot exploit reveals security nightmare for
  Android}.
\newblock
  \bibinfo{howpublished}{https://geeksided.com/2014/06/16/towelroot-exploit-reveals-security-nightmare-android/}.
\newblock


\bibitem[\protect\citeauthoryear{??}{cbe}{2016}]%
        {cbenchmarks}
 \bibinfo{year}{2016}\natexlab{}.
\newblock \bibinfo{title}{tum-i22/obfuscation-benchmarks}.
\newblock
  \bibinfo{howpublished}{https://github.com/tum-i22/obfuscation-benchmarks}.
\newblock


\bibitem[\protect\citeauthoryear{??}{mal}{2017a}]%
        {malware2}
 \bibinfo{year}{2017}\natexlab{a}.
\newblock \bibinfo{title}{Deep Analysis of Android Rootnik Malware Using
  Advanced Anti-Debug and Anti-Hook, Part I: Debugging in The Scope of Native
  Layer}.
\newblock
  \bibinfo{howpublished}{https://www.fortinet.com/blog/threat-research/deep-analysis-of-android-rootnik-malware-using-advanced-anti-debug-and-anti-hook-part-i-debugging-in-the-scope-of-native-layer.html}.
\newblock


\bibitem[\protect\citeauthoryear{??}{mal}{2017b}]%
        {malware3}
 \bibinfo{year}{2017}\natexlab{b}.
\newblock \bibinfo{title}{ZNIU: First Android Malware to Exploit Dirty COW
  Vulnerability}.
\newblock
  \bibinfo{howpublished}{https://blog.trendmicro.com/trendlabs-security-intelligence/zniu-first-android-malware-exploit-dirty-cow-vulnerability/}.
\newblock


\bibitem[\protect\citeauthoryear{??}{all}{2018}]%
        {allatori}
 \bibinfo{year}{2018}\natexlab{}.
\newblock \bibinfo{title}{Allatori Java Obfuscator - Professional Java
  Obfuscation}.
\newblock \bibinfo{howpublished}{www.allatori.com/}.
\newblock


\bibitem[\protect\citeauthoryear{??}{NDK}{2018}]%
        {NDK}
 \bibinfo{year}{2018}\natexlab{}.
\newblock \bibinfo{title}{Android NDK - Android Developers}.
\newblock \bibinfo{howpublished}{https://developer.android.com/ndk/}.
\newblock


\bibitem[\protect\citeauthoryear{??}{Dex}{2018a}]%
        {DexGuard}
 \bibinfo{year}{2018}\natexlab{a}.
\newblock \bibinfo{title}{Android obfuscation and runtime-self protection
  (RASP) | DexGuard}.
\newblock
  \bibinfo{howpublished}{https://www.guardsquare.com/en/products/dexguard}.
\newblock


\bibitem[\protect\citeauthoryear{??}{bai}{2018}]%
        {baidujiagu}
 \bibinfo{year}{2018}\natexlab{}.
\newblock \bibinfo{title}{Baidu Jiagu}.
\newblock \bibinfo{howpublished}{https://app.baidu.com/jiagu/}.
\newblock


\bibitem[\protect\citeauthoryear{??}{Ban}{2018}]%
        {Bangcle}
 \bibinfo{year}{2018}\natexlab{}.
\newblock \bibinfo{title}{Bangcle Jiagu}.
\newblock \bibinfo{howpublished}{https://www.bangcle.com/}.
\newblock


\bibitem[\protect\citeauthoryear{??}{bin}{2018}]%
        {binaryninja}
 \bibinfo{year}{2018}\natexlab{}.
\newblock \bibinfo{title}{binary.ninja : a reverse engineering platform}.
\newblock \bibinfo{howpublished}{https://binary.ninja/}.
\newblock


\bibitem[\protect\citeauthoryear{??}{Dex}{2018b}]%
        {DexProtector}
 \bibinfo{year}{2018}\natexlab{b}.
\newblock \bibinfo{title}{DexProtector by Licel — Cutting edge obfuscator for
  Android apps}.
\newblock \bibinfo{howpublished}{https://dexprotector.com/}.
\newblock


\bibitem[\protect\citeauthoryear{??}{D-O}{2018}]%
        {D-OLLVM2017}
 \bibinfo{year}{2018}\natexlab{}.
\newblock \bibinfo{title}{Dissecting LLVM Obfuscator Part 1}.
\newblock
  \bibinfo{howpublished}{https://blog.rpis.ec/2017/12/dissection-llvm-obfuscator-p1.html}.
\newblock


\bibitem[\protect\citeauthoryear{??}{F-D}{2018}]%
        {F-Droid}
 \bibinfo{year}{2018}\natexlab{}.
\newblock \bibinfo{title}{F-Droid - Free and Open Source Android App
  Repository}.
\newblock \bibinfo{howpublished}{https://f-droid.org/zh\_Hant/about/}.
\newblock


\bibitem[\protect\citeauthoryear{??}{mia}{2018}]%
        {miasm}
 \bibinfo{year}{2018}\natexlab{}.
\newblock \bibinfo{title}{GitHub - cea-sec/miasm: Reverse engineering framework
  in Python}.
\newblock \bibinfo{howpublished}{https://github.com/cea-sec/miasm}.
\newblock


\bibitem[\protect\citeauthoryear{??}{mal}{2018}]%
        {malware1}
 \bibinfo{year}{2018}\natexlab{}.
\newblock \bibinfo{title}{Google Researcher Unpacks Rare Android Malware
  Obfuscation Library}.
\newblock
  \bibinfo{howpublished}{https://www.darkreading.com/attacks-breaches/google-researcher-unpacks-rare-android-malware-obfuscation-library-/d/d-id/1332444}.
\newblock


\bibitem[\protect\citeauthoryear{??}{Das}{2018}]%
        {DashO}
 \bibinfo{year}{2018}\natexlab{}.
\newblock \bibinfo{title}{Java Obfuscator \& Android Obfuscator | DashO -
  PreEmptive Solutions}.
\newblock
  \bibinfo{howpublished}{https://www.preemptive.com/products/dasho/overview}.
\newblock


\bibitem[\protect\citeauthoryear{??}{pro}{2018}]%
        {proguard}
 \bibinfo{year}{2018}\natexlab{}.
\newblock \bibinfo{title}{ProGuard}.
\newblock
  \bibinfo{howpublished}{https://www.guardsquare.com/en/products/proguard}.
\newblock


\bibitem[\protect\citeauthoryear{Afonso, Bianchi, Fratantonio, Doup{\'e},
  Polino, de~Geus, Kruegel, and Vigna}{Afonso et~al\mbox{.}}{2016}]%
        {afonso2016going}
\bibfield{author}{\bibinfo{person}{Vitor Afonso}, \bibinfo{person}{Antonio
  Bianchi}, \bibinfo{person}{Yanick Fratantonio}, \bibinfo{person}{Adam
  Doup{\'e}}, \bibinfo{person}{Mario Polino}, \bibinfo{person}{Paulo de Geus},
  \bibinfo{person}{Christopher Kruegel}, {and} \bibinfo{person}{Giovanni
  Vigna}.} \bibinfo{year}{2016}\natexlab{}.
\newblock \showarticletitle{Going native: Using a large-scale analysis of
  android apps to create a practical native-code sandboxing policy}. In
  \bibinfo{booktitle}{\emph{The Network and Distributed System Security
  Symposium}}. \bibinfo{pages}{1--15}.
\newblock


\bibitem[\protect\citeauthoryear{Alam, Qu, Riley, Chen, and Rastogi}{Alam
  et~al\mbox{.}}{2016}]%
        {alam2016droidnative}
\bibfield{author}{\bibinfo{person}{Shahid Alam}, \bibinfo{person}{Zhengyang
  Qu}, \bibinfo{person}{Ryan Riley}, \bibinfo{person}{Yan Chen}, {and}
  \bibinfo{person}{Vaibhav Rastogi}.} \bibinfo{year}{2016}\natexlab{}.
\newblock \showarticletitle{Droidnative: Semantic-based detection of android
  native code malware}.
\newblock \bibinfo{journal}{\emph{arXiv preprint arXiv:1602.04693}}
  (\bibinfo{year}{2016}).
\newblock


\bibitem[\protect\citeauthoryear{Anand, P{\u{a}}s{\u{a}}reanu, and
  Visser}{Anand et~al\mbox{.}}{2007}]%
        {anand2007jpf}
\bibfield{author}{\bibinfo{person}{Saswat Anand}, \bibinfo{person}{Corina~S
  P{\u{a}}s{\u{a}}reanu}, {and} \bibinfo{person}{Willem Visser}.}
  \bibinfo{year}{2007}\natexlab{}.
\newblock \showarticletitle{JPF--SE: A symbolic execution extension to java
  pathfinder}. In \bibinfo{booktitle}{\emph{International Conference on Tools
  and Algorithms for the Construction and Analysis of Systems}}. Springer,
  \bibinfo{pages}{134--138}.
\newblock


\bibitem[\protect\citeauthoryear{AppBrain}{AppBrain}{2018}]%
        {AppNumber}
\bibfield{author}{\bibinfo{person}{AppBrain}.} \bibinfo{year}{2018}\natexlab{}.
\newblock \bibinfo{title}{Current number of Android apps on Google Play}.
\newblock \bibinfo{howpublished}{https://www.appbrain.com/stats}.
\newblock


\bibitem[\protect\citeauthoryear{Banescu, Collberg, Ganesh, Newsham, and
  Pretschner}{Banescu et~al\mbox{.}}{2016}]%
        {banescu2016code}
\bibfield{author}{\bibinfo{person}{Sebastian Banescu},
  \bibinfo{person}{Christian Collberg}, \bibinfo{person}{Vijay Ganesh},
  \bibinfo{person}{Zack Newsham}, {and} \bibinfo{person}{Alexander
  Pretschner}.} \bibinfo{year}{2016}\natexlab{}.
\newblock \showarticletitle{Code obfuscation against symbolic execution
  attacks}. In \bibinfo{booktitle}{\emph{Proceedings of the 32nd Annual
  Conference on Computer Security Applications}}. ACM,
  \bibinfo{pages}{189--200}.
\newblock


\bibitem[\protect\citeauthoryear{Baumann, Protsenko, and Müller}{Baumann
  et~al\mbox{.}}{2017}]%
        {Baumann2017Anti}
\bibfield{author}{\bibinfo{person}{Richard Baumann}, \bibinfo{person}{Mykolai
  Protsenko}, {and} \bibinfo{person}{Tilo Müller}.}
  \bibinfo{year}{2017}\natexlab{}.
\newblock \showarticletitle{Anti-ProGuard: Towards Automated Deobfuscation of
  Android Apps}. In \bibinfo{booktitle}{\emph{The Workshop on Security in
  Highly Connected It Systems}}. \bibinfo{pages}{7--12}.
\newblock


\bibitem[\protect\citeauthoryear{Bichsel, Raychev, Tsankov, and Vechev}{Bichsel
  et~al\mbox{.}}{2016}]%
        {Bichsel2016Statistical}
\bibfield{author}{\bibinfo{person}{Benjamin Bichsel}, \bibinfo{person}{Veselin
  Raychev}, \bibinfo{person}{Petar Tsankov}, {and} \bibinfo{person}{Martin
  Vechev}.} \bibinfo{year}{2016}\natexlab{}.
\newblock \showarticletitle{Statistical Deobfuscation of Android Applications}.
  In \bibinfo{booktitle}{\emph{ACM Sigsac Conference on Computer and
  Communications Security}}. \bibinfo{pages}{343--355}.
\newblock


\bibitem[\protect\citeauthoryear{{Bird}}{{Bird}}{2017}]%
        {CFFRef}
\bibfield{author}{\bibinfo{person}{{Bird}}.} \bibinfo{year}{2017}\natexlab{}.
\newblock \bibinfo{title}{Symbolic Execution based approach to remove control
  flow flattening}.
\newblock
  \bibinfo{howpublished}{https://security.tencent.com/index.php/blog/msg/112}.
\newblock


\bibitem[\protect\citeauthoryear{Cadar, Godefroid, Khurshid, Pasareanu, Sen,
  Tillmann, and Visser}{Cadar et~al\mbox{.}}{2011}]%
        {cadar2011symbolic}
\bibfield{author}{\bibinfo{person}{Cristian Cadar}, \bibinfo{person}{Patrice
  Godefroid}, \bibinfo{person}{Sarfraz Khurshid}, \bibinfo{person}{Corina~S
  Pasareanu}, \bibinfo{person}{Koushik Sen}, \bibinfo{person}{Nikolai
  Tillmann}, {and} \bibinfo{person}{Willem Visser}.}
  \bibinfo{year}{2011}\natexlab{}.
\newblock \showarticletitle{Symbolic execution for software testing in
  practice: preliminary assessment}. In \bibinfo{booktitle}{\emph{Software
  Engineering (ICSE), 2011 33rd International Conference on}}. IEEE,
  \bibinfo{pages}{1066--1071}.
\newblock


\bibitem[\protect\citeauthoryear{Cha, Avgerinos, Rebert, and Brumley}{Cha
  et~al\mbox{.}}{2012}]%
        {cha2012unleashing}
\bibfield{author}{\bibinfo{person}{Sang~Kil Cha}, \bibinfo{person}{Thanassis
  Avgerinos}, \bibinfo{person}{Alexandre Rebert}, {and} \bibinfo{person}{David
  Brumley}.} \bibinfo{year}{2012}\natexlab{}.
\newblock \showarticletitle{Unleashing mayhem on binary code}. In
  \bibinfo{booktitle}{\emph{Security and Privacy (SP), 2012 IEEE Symposium
  on}}. IEEE, \bibinfo{pages}{380--394}.
\newblock


\bibitem[\protect\citeauthoryear{Chan and Collberg}{Chan and Collberg}{2014}]%
        {chan2014method}
\bibfield{author}{\bibinfo{person}{Patrick~PF Chan} {and}
  \bibinfo{person}{Christian Collberg}.} \bibinfo{year}{2014}\natexlab{}.
\newblock \showarticletitle{A method to evaluate CFG comparison algorithms}. In
  \bibinfo{booktitle}{\emph{Quality Software (QSIC), 2014 14th International
  Conference on}}. IEEE, \bibinfo{pages}{95--104}.
\newblock


\bibitem[\protect\citeauthoryear{David}{David}{2017}]%
        {david2017formal}
\bibfield{author}{\bibinfo{person}{Robin David}.}
  \bibinfo{year}{2017}\natexlab{}.
\newblock \emph{\bibinfo{title}{Formal Approaches for Automatic Deobfuscation
  and Reverse-engineering of Protected Codes}}.
\newblock \bibinfo{thesistype}{Ph.D. Dissertation}.
  \bibinfo{school}{Universit{\'e} de Lorraine}.
\newblock


\bibitem[\protect\citeauthoryear{{Decompiler, JEB}}{{Decompiler, JEB}}{2017}]%
        {decompiler_2017}
\bibfield{author}{\bibinfo{person}{{Decompiler, JEB}}.}
  \bibinfo{year}{2017}\natexlab{}.
\newblock \bibinfo{title}{JEB 2.3 Native Decompilation, Advanced Deobfuscation
  demo part 2 (CFG Unflattening on x86), video+text here
  https://youtu.be/jIcz74NT2Pg}.
\newblock
  \bibinfo{howpublished}{https://twitter.com/jebdec/status/832735845488537600}.
\newblock


\bibitem[\protect\citeauthoryear{{Decompiler, JEB}}{{Decompiler, JEB}}{2020}]%
        {jeb}
\bibfield{author}{\bibinfo{person}{{Decompiler, JEB}}.}
  \bibinfo{year}{2020}\natexlab{}.
\newblock \bibinfo{title}{JEB by PNF Software}.
\newblock \bibinfo{howpublished}{https://www.pnfsoftware.com/}.
\newblock


\bibitem[\protect\citeauthoryear{Dong, Li, Diao, Liu, Liu, Li, Xu, Chen, Wang,
  and Zhang}{Dong et~al\mbox{.}}{2018}]%
        {dong2018understanding}
\bibfield{author}{\bibinfo{person}{Shuaike Dong}, \bibinfo{person}{Menghao Li},
  \bibinfo{person}{Wenrui Diao}, \bibinfo{person}{Xiangyu Liu},
  \bibinfo{person}{Jian Liu}, \bibinfo{person}{Zhou Li},
  \bibinfo{person}{Fenghao Xu}, \bibinfo{person}{Kai Chen},
  \bibinfo{person}{Xiaofeng Wang}, {and} \bibinfo{person}{Kehuan Zhang}.}
  \bibinfo{year}{2018}\natexlab{}.
\newblock \showarticletitle{Understanding Android Obfuscation Techniques: A
  Large-Scale Investigation in the Wild}.
\newblock \bibinfo{journal}{\emph{arXiv preprint arXiv:1801.01633}}
  (\bibinfo{year}{2018}).
\newblock


\bibitem[\protect\citeauthoryear{Duan, Zhang, Bhaskar, Yin, Pan, Li, Wang, and
  Wang}{Duan et~al\mbox{.}}{2018}]%
        {duan2018things}
\bibfield{author}{\bibinfo{person}{Yue Duan}, \bibinfo{person}{Mu Zhang},
  \bibinfo{person}{Abhishek~Vasisht Bhaskar}, \bibinfo{person}{Heng Yin},
  \bibinfo{person}{Xiaorui Pan}, \bibinfo{person}{Tongxin Li},
  \bibinfo{person}{Xueqiang Wang}, {and} \bibinfo{person}{X Wang}.}
  \bibinfo{year}{2018}\natexlab{}.
\newblock \showarticletitle{Things you may not know about android (un) packers:
  a systematic study based on whole-system emulation}. In
  \bibinfo{booktitle}{\emph{25th Annual Network and Distributed System Security
  Symposium, NDSS}}. \bibinfo{pages}{18--21}.
\newblock


\bibitem[\protect\citeauthoryear{Fyrbiak, Rokicki, Bissantz, Tessier, and
  Paar}{Fyrbiak et~al\mbox{.}}{2018}]%
        {fyrbiak2018hybrid}
\bibfield{author}{\bibinfo{person}{Marc Fyrbiak}, \bibinfo{person}{Simon
  Rokicki}, \bibinfo{person}{Nicolai Bissantz}, \bibinfo{person}{Russell
  Tessier}, {and} \bibinfo{person}{Christof Paar}.}
  \bibinfo{year}{2018}\natexlab{}.
\newblock \showarticletitle{Hybrid Obfuscation to Protect Against Disclosure
  Attacks on Embedded Microprocessors}.
\newblock \bibinfo{journal}{\emph{IEEE Trans. Comput.}} \bibinfo{volume}{67},
  \bibinfo{number}{3} (\bibinfo{year}{2018}), \bibinfo{pages}{307--321}.
\newblock


\bibitem[\protect\citeauthoryear{Gabriel}{Gabriel}{2014}]%
        {gabriel2014deobfuscation}
\bibfield{author}{\bibinfo{person}{F Gabriel}.}
  \bibinfo{year}{2014}\natexlab{}.
\newblock \showarticletitle{Deobfuscation: recovering an OLLVM-protected
  program}.
\newblock \bibinfo{journal}{\emph{QuarkLabs}}  \bibinfo{volume}{4}
  (\bibinfo{year}{2014}), \bibinfo{pages}{12}.
\newblock


\bibitem[\protect\citeauthoryear{Guillot and Gazet}{Guillot and Gazet}{2010}]%
        {guillot2010automatic}
\bibfield{author}{\bibinfo{person}{Yoann Guillot} {and}
  \bibinfo{person}{Alexandre Gazet}.} \bibinfo{year}{2010}\natexlab{}.
\newblock \showarticletitle{Automatic binary deobfuscation}.
\newblock \bibinfo{journal}{\emph{Journal in computer virology}}
  \bibinfo{volume}{6}, \bibinfo{number}{3} (\bibinfo{year}{2010}),
  \bibinfo{pages}{261--276}.
\newblock


\bibitem[\protect\citeauthoryear{Hu, Chiueh, and Shin}{Hu
  et~al\mbox{.}}{2009}]%
        {hu2009large}
\bibfield{author}{\bibinfo{person}{Xin Hu}, \bibinfo{person}{Tzi-cker Chiueh},
  {and} \bibinfo{person}{Kang~G Shin}.} \bibinfo{year}{2009}\natexlab{}.
\newblock \showarticletitle{Large-scale malware indexing using function-call
  graphs}. In \bibinfo{booktitle}{\emph{Proceedings of the 16th ACM conference
  on Computer and communications security}}. ACM, \bibinfo{pages}{611--620}.
\newblock


\bibitem[\protect\citeauthoryear{Junod, Rinaldini, Wehrli, and Michielin}{Junod
  et~al\mbox{.}}{2015}]%
        {junod2015obfuscator}
\bibfield{author}{\bibinfo{person}{Pascal Junod}, \bibinfo{person}{Julien
  Rinaldini}, \bibinfo{person}{Johan Wehrli}, {and} \bibinfo{person}{Julie
  Michielin}.} \bibinfo{year}{2015}\natexlab{}.
\newblock \showarticletitle{Obfuscator-LLVM--software protection for the
  masses}. In \bibinfo{booktitle}{\emph{Software Protection (SPRO), 2015
  IEEE/ACM 1st International Workshop on}}. IEEE, \bibinfo{pages}{3--9}.
\newblock


\bibitem[\protect\citeauthoryear{Kotov and Wojnowicz}{Kotov and
  Wojnowicz}{2018}]%
        {kotov2018towards}
\bibfield{author}{\bibinfo{person}{Vadim Kotov} {and} \bibinfo{person}{Michael
  Wojnowicz}.} \bibinfo{year}{2018}\natexlab{}.
\newblock \showarticletitle{Towards Generic Deobfuscation of Windows API
  Calls}.
\newblock \bibinfo{journal}{\emph{arXiv preprint arXiv:1802.04466}}
  (\bibinfo{year}{2018}).
\newblock


\bibitem[\protect\citeauthoryear{Kuhn}{Kuhn}{1955}]%
        {kuhn1955hungarian}
\bibfield{author}{\bibinfo{person}{Harold~W Kuhn}.}
  \bibinfo{year}{1955}\natexlab{}.
\newblock \showarticletitle{The Hungarian method for the assignment problem}.
\newblock \bibinfo{journal}{\emph{Naval research logistics quarterly}}
  \bibinfo{volume}{2}, \bibinfo{number}{1-2} (\bibinfo{year}{1955}),
  \bibinfo{pages}{83--97}.
\newblock


\bibitem[\protect\citeauthoryear{Likarish, Jung, and Jo}{Likarish
  et~al\mbox{.}}{2009}]%
        {likarish2009obfuscated}
\bibfield{author}{\bibinfo{person}{Peter Likarish}, \bibinfo{person}{Eunjin
  Jung}, {and} \bibinfo{person}{Insoon Jo}.} \bibinfo{year}{2009}\natexlab{}.
\newblock \showarticletitle{Obfuscated malicious javascript detection using
  classification techniques}. In \bibinfo{booktitle}{\emph{Malicious and
  Unwanted Software (MALWARE), 2009 4th International Conference on}}. IEEE,
  \bibinfo{pages}{47--54}.
\newblock


\bibitem[\protect\citeauthoryear{{liumengdeqq}}{{liumengdeqq}}{2018}]%
        {CFFRef2}
\bibfield{author}{\bibinfo{person}{{liumengdeqq}}.}
  \bibinfo{year}{2018}\natexlab{}.
\newblock \bibinfo{title}{control flow flattening}.
\newblock \bibinfo{howpublished}{https://github.com/liumengdeqq/deflat}.
\newblock


\bibitem[\protect\citeauthoryear{Madou, Van~Put, and De~Bosschere}{Madou
  et~al\mbox{.}}{2006}]%
        {madou2006loco}
\bibfield{author}{\bibinfo{person}{Matias Madou}, \bibinfo{person}{Ludo
  Van~Put}, {and} \bibinfo{person}{Koen De~Bosschere}.}
  \bibinfo{year}{2006}\natexlab{}.
\newblock \showarticletitle{Loco: An interactive code (de) obfuscation tool}.
  In \bibinfo{booktitle}{\emph{Proceedings of the 2006 ACM SIGPLAN symposium on
  Partial evaluation and semantics-based program manipulation}}. ACM,
  \bibinfo{pages}{140--144}.
\newblock


\bibitem[\protect\citeauthoryear{Pawlowski, Contag, and Holz}{Pawlowski
  et~al\mbox{.}}{2016}]%
        {pawlowski2016probfuscation}
\bibfield{author}{\bibinfo{person}{Andre Pawlowski}, \bibinfo{person}{Moritz
  Contag}, {and} \bibinfo{person}{Thorsten Holz}.}
  \bibinfo{year}{2016}\natexlab{}.
\newblock \showarticletitle{Probfuscation: an obfuscation approach using
  probabilistic control flows}. In \bibinfo{booktitle}{\emph{International
  Conference on Detection of Intrusions and Malware, and Vulnerability
  Assessment}}. Springer, \bibinfo{pages}{165--185}.
\newblock


\bibitem[\protect\citeauthoryear{Pearson}{Pearson}{1896}]%
        {pearson1896mathematical}
\bibfield{author}{\bibinfo{person}{Karl Pearson}.}
  \bibinfo{year}{1896}\natexlab{}.
\newblock \showarticletitle{Mathematical contributions to the theory of
  evolution. III. Regression, heredity, and panmixia}.
\newblock \bibinfo{journal}{\emph{Philosophical Transactions of the Royal
  Society of London. Series A, containing papers of a mathematical or physical
  character}}  \bibinfo{volume}{187} (\bibinfo{year}{1896}),
  \bibinfo{pages}{253--318}.
\newblock


\bibitem[\protect\citeauthoryear{Salwan, Bardin, and Potet}{Salwan
  et~al\mbox{.}}{2018}]%
        {salwan2018symbolic}
\bibfield{author}{\bibinfo{person}{Jonathan Salwan},
  \bibinfo{person}{S{\'e}bastien Bardin}, {and} \bibinfo{person}{Marie-Laure
  Potet}.} \bibinfo{year}{2018}\natexlab{}.
\newblock \showarticletitle{Symbolic deobfuscation: From virtualized code back
  to the original}. In \bibinfo{booktitle}{\emph{International Conference on
  Detection of Intrusions and Malware, and Vulnerability Assessment}}.
  Springer, \bibinfo{pages}{372--392}.
\newblock


\bibitem[\protect\citeauthoryear{Shoshitaishvili, Wang, Salls, Stephens,
  Polino, Dutcher, Grosen, Feng, Hauser, Kruegel, and Vigna}{Shoshitaishvili
  et~al\mbox{.}}{2016}]%
        {shoshitaishvili2016state}
\bibfield{author}{\bibinfo{person}{Yan Shoshitaishvili}, \bibinfo{person}{Ruoyu
  Wang}, \bibinfo{person}{Christopher Salls}, \bibinfo{person}{Nick Stephens},
  \bibinfo{person}{Mario Polino}, \bibinfo{person}{Audrey Dutcher},
  \bibinfo{person}{John Grosen}, \bibinfo{person}{Siji Feng},
  \bibinfo{person}{Christophe Hauser}, \bibinfo{person}{Christopher Kruegel},
  {and} \bibinfo{person}{Giovanni Vigna}.} \bibinfo{year}{2016}\natexlab{}.
\newblock \showarticletitle{{SoK: (State of) The Art of War: Offensive
  Techniques in Binary Analysis}}. In \bibinfo{booktitle}{\emph{IEEE Symposium
  on Security and Privacy}}.
\newblock


\bibitem[\protect\citeauthoryear{Sun and Tan}{Sun and Tan}{2014}]%
        {sun2014nativeguard}
\bibfield{author}{\bibinfo{person}{Mengtao Sun} {and} \bibinfo{person}{Gang
  Tan}.} \bibinfo{year}{2014}\natexlab{}.
\newblock \showarticletitle{Nativeguard: Protecting android applications from
  third-party native libraries}. In \bibinfo{booktitle}{\emph{Proceedings of
  the 2014 ACM conference on Security and privacy in wireless \& mobile
  networks}}. ACM, \bibinfo{pages}{165--176}.
\newblock


\bibitem[\protect\citeauthoryear{Trabish, Mattavelli, Rinetzky, and
  Cadar}{Trabish et~al\mbox{.}}{2018}]%
        {trabish2018chopped}
\bibfield{author}{\bibinfo{person}{David Trabish}, \bibinfo{person}{Andrea
  Mattavelli}, \bibinfo{person}{Noam Rinetzky}, {and} \bibinfo{person}{Cristian
  Cadar}.} \bibinfo{year}{2018}\natexlab{}.
\newblock \showarticletitle{Chopped symbolic execution}. In
  \bibinfo{booktitle}{\emph{Proceedings of the 40th International Conference on
  Software Engineering}}. ACM, \bibinfo{pages}{350--360}.
\newblock


\bibitem[\protect\citeauthoryear{Udupa, Debray, and Madou}{Udupa
  et~al\mbox{.}}{2005}]%
        {udupa2005deobfuscation}
\bibfield{author}{\bibinfo{person}{Sharath~K Udupa}, \bibinfo{person}{Saumya~K
  Debray}, {and} \bibinfo{person}{Matias Madou}.}
  \bibinfo{year}{2005}\natexlab{}.
\newblock \showarticletitle{Deobfuscation: Reverse engineering obfuscated
  code}. In \bibinfo{booktitle}{\emph{Reverse Engineering, 12th Working
  Conference on}}. IEEE, \bibinfo{pages}{10--pp}.
\newblock


\bibitem[\protect\citeauthoryear{{Wikipedia contributors}}{{Wikipedia
  contributors}}{2016}]%
        {wiki:opaque}
\bibfield{author}{\bibinfo{person}{{Wikipedia contributors}}.}
  \bibinfo{year}{2016}\natexlab{}.
\newblock \bibinfo{title}{Opaque predicate --- {Wikipedia}{,} The Free
  Encyclopedia}.
\newblock
  \bibinfo{howpublished}{https://en.wikipedia.org/w/index.php?\\title=Opaque\_predicate\&oldid=741606824}.
\newblock
\newblock
\shownote{[Online; accessed 25-August-2018].}


\bibitem[\protect\citeauthoryear{{Wikipedia contributors}}{{Wikipedia
  contributors}}{2018a}]%
        {wiki:confold}
\bibfield{author}{\bibinfo{person}{{Wikipedia contributors}}.}
  \bibinfo{year}{2018}\natexlab{a}.
\newblock \bibinfo{title}{Constant folding --- {Wikipedia}{,} The Free
  Encyclopedia}.
\newblock
\newblock
\urldef\tempurl%
\url{https://en.wikipedia.org/w/index.php?\\title=Constant_folding&oldid=855984743}
\showURL{%
\tempurl}
\newblock
\shownote{[Online; accessed 25-August-2018].}


\bibitem[\protect\citeauthoryear{{Wikipedia contributors}}{{Wikipedia
  contributors}}{2018b}]%
        {wiki:distance}
\bibfield{author}{\bibinfo{person}{{Wikipedia contributors}}.}
  \bibinfo{year}{2018}\natexlab{b}.
\newblock \bibinfo{title}{Euclidean distance --- {Wikipedia}{,} The Free
  Encyclopedia}.
\newblock
  \bibinfo{howpublished}{https://en.wikipedia.org/w/index.php?\\title=Euclidean\_distance\&oldid=855266282}.
\newblock
\newblock
\shownote{[Online; accessed 25-August-2018].}


\bibitem[\protect\citeauthoryear{{Wikipedia contributors}}{{Wikipedia
  contributors}}{2018c}]%
        {wiki:Library}
\bibfield{author}{\bibinfo{person}{{Wikipedia contributors}}.}
  \bibinfo{year}{2018}\natexlab{c}.
\newblock \bibinfo{title}{Library (computing) --- {Wikipedia}{,} The Free
  Encyclopedia}.
\newblock
\newblock
\urldef\tempurl%
\url{https://en.wikipedia.org/w/index.php?title=Library_(computing)&oldid=852686325}
\showURL{%
\tempurl}
\newblock
\shownote{[Online; accessed 22-August-2018].}


\bibitem[\protect\citeauthoryear{{Wikipedia contributors}}{{Wikipedia
  contributors}}{2018d}]%
        {wiki:symbexe}
\bibfield{author}{\bibinfo{person}{{Wikipedia contributors}}.}
  \bibinfo{year}{2018}\natexlab{d}.
\newblock \bibinfo{title}{Symbolic execution --- {Wikipedia}{,} The Free
  Encyclopedia}.
\newblock
\newblock
\urldef\tempurl%
\url{https://en.wikipedia.org/w/index.php?title=Symbolic_execution&oldid=835809262}
\showURL{%
\tempurl}
\newblock
\shownote{[Online; accessed 23-August-2018].}


\bibitem[\protect\citeauthoryear{Xu, Ming, and Wu}{Xu et~al\mbox{.}}{2017}]%
        {xu2017cryptographic}
\bibfield{author}{\bibinfo{person}{Dongpeng Xu}, \bibinfo{person}{Jiang Ming},
  {and} \bibinfo{person}{Dinghao Wu}.} \bibinfo{year}{2017}\natexlab{}.
\newblock \showarticletitle{Cryptographic function detection in obfuscated
  binaries via bit-precise symbolic loop mapping}. In
  \bibinfo{booktitle}{\emph{Security and Privacy (SP), 2017 IEEE Symposium
  on}}. IEEE, \bibinfo{pages}{921--937}.
\newblock


\bibitem[\protect\citeauthoryear{Yadegari}{Yadegari}{2016}]%
        {yadegari2016automatic}
\bibfield{author}{\bibinfo{person}{Babak Yadegari}.}
  \bibinfo{year}{2016}\natexlab{}.
\newblock \showarticletitle{Automatic deobfuscation and reverse engineering of
  obfuscated code}.
\newblock  (\bibinfo{year}{2016}).
\newblock


\bibitem[\protect\citeauthoryear{Yadegari, Johannesmeyer, Whitely, and
  Debray}{Yadegari et~al\mbox{.}}{2015}]%
        {yadegari2015generic}
\bibfield{author}{\bibinfo{person}{Babak Yadegari}, \bibinfo{person}{Brian
  Johannesmeyer}, \bibinfo{person}{Ben Whitely}, {and} \bibinfo{person}{Saumya
  Debray}.} \bibinfo{year}{2015}\natexlab{}.
\newblock \showarticletitle{A generic approach to automatic deobfuscation of
  executable code}. In \bibinfo{booktitle}{\emph{2015 IEEE Symposium on
  Security and Privacy (SP)}}. IEEE, \bibinfo{pages}{674--691}.
\newblock


\end{thebibliography}

\end{document}